\documentclass[aps,pre,reprint,superscriptaddress,amsmath,amssymb,showpacs,floatfix,twocolumn]{revtex4-1}
\usepackage{graphicx}
\usepackage{bm}
\usepackage{color}
\usepackage{soul}
\usepackage{dcolumn}
\usepackage{multirow}
\usepackage{tikz}
\usepackage{hhline}
\usepackage{mathtools}
\usepackage{verbatim}

\newcommand{\ave}[1]{\langle #1 \rangle}

\newcommand{\be}{\begin{equation}}
\newcommand{\ee}{\end{equation}}
\newcommand{\ba}{\begin{eqnarray}}
\newcommand{\ea}{\end{eqnarray}}

\definecolor{asparagus}{rgb}{0.53, 0.66, 0.42}
\definecolor{cadmiumgreen}{rgb}{0.0, 0.42, 0.24}

\DeclareMathOperator{\sech}{sech}

\begin{document}
\title{Estimating Free Energy Differences with Virtually Escorted Trajectories}

\author{Sangyun Lee}
\affiliation{Institut für Physik, Johannes Gutenberg University Mainz,
Staudingerweg 9, 55128 Mainz, Germany.}
\affiliation{Institute for Physical Science and Technology, University of Maryland, College Park, Maryland 20742, USA}
\affiliation{School of Physics, Korea Institute for Advanced Study, Seoul, 02455, Korea}
\author{Christopher Jarzynski}
\email{cjarzyns@umd.edu}
\affiliation{Institute for Physical Science and Technology, University of Maryland, College Park, Maryland 20742, USA}
\affiliation{Department of Chemistry and Biochemistry, University of Maryland, College Park, MD 20742 USA}
\affiliation{Department of Physics, University of Maryland, College Park, MD 20742 USA}

\date{\today}

\begin{abstract}
For a process in which a system is driven irreversibly from equilibrium state $A$ toward equilibrium state $B$,
the free energy difference $\Delta F = F_B-F_A$ can be estimated using the work fluctuation theorem $\langle e^{-W/T}\rangle = e^{-\Delta F/T}$, where $W$ and $T$ denote work and temperature.
The estimate often suffers from poor convergence with the number of trajectories used to calculate the average.
Borrowing ideas from escorted free energy estimation, and from diffusion models of machine learning, we show how to construct infinitely many work-like quantities, $W_\theta$, that satisfy $\langle e^{-W_\theta/T}\rangle = e^{-\Delta F/T}$, for the same underlying dynamics.
Our method involves a virtual control field ${\bm u}_\theta$ that does not modify these dynamics.
We show how to choose parameter values $\theta$ to optimize convergence of the free energy estimate, for a fixed set of trajectories.
We identify conditions under which our method provides a zero-variance estimator of $\Delta F$.
We use numerical simulations of model systems to illustrate the gains in convergence that our method can achieve.
\end{abstract}

\maketitle

{\it Introduction---}The work fluctuation theorem~\cite{Chris1997Nonequilibrium}, 
\be
\label{eq:wft}
\left\langle e^{-W/T} \right\rangle = e^{-\Delta F/T} \, ,
\ee
relates equilibrium free energy differences to nonequilibrium work fluctuations.
$\Delta F$ is the free energy difference between a system's equilibrium states, $A$ and $B$; $T$ is the temperature of both states; $W$ is the work performed on the system during one realization of a finite-time process that drives it from $A$ toward $B$; angular brackets denote an ensemble average over realizations; and we set Boltzmann's constant $k_B$ to unity.
In the context of free energy estimation, the left side of Eq.~\eqref{eq:wft} can be evaluated by averaging over a finite number, $N$, of simulated or experimentally measured realizations.
However, the convergence of the average with $N$ is often slow, resulting in a poor estimate of $\Delta F$~\cite{ChrisPRE1997}.

Convergence can be improved by adding an artificial control field, ${\bm u}$, to the system's dynamics (see {\it Background}).
If the control field significantly reduces the degree to which the system lags~\cite{vaikuntanathan2009dissipation} behind the instantaneous equilibrium state, then the convergence of the estimate of $\Delta F$ can be accelerated dramatically~\cite{escortedPRL2008}.
This approach is known as {\it escorted} free energy estimation, as it relies on the design of a field ${\bm u}$ that ``escorts'' the system along a near-equilibrium path.
One of the limitations of the method is that the design of effective control fields is challenging, though machine learning-based approaches can help in this regard~\cite{Wirnsberger2020TargetedML,vargas2024transport,Zhong2024TimeAsymmetric,Mate2024Neural,Mate2025,rosa2025nonadiabatic,he2025feat,Holdijk2026,Schebek2026}.
Moreover, iterative improvements in ${\bm u}$ require new simulations. Lastly, the method is not easily extended to the experimental context.

Here, we introduce a method that delivers the benefits of escorted free energy estimation, without requiring an explicit control field.
The key idea is to apply {\it virtual} control at the post-processing level. 
For a given process from state $A$ toward $B$, and for trajectories generated {\it without} a control field, we show that there exist infinitely many work-like quantities $W_\theta$ that satisfy the identity
\be
\label{eq:ewft-intro}
\left\langle e^{-W_\theta/T} \right\rangle = e^{-\Delta F/T} \, .
\ee
Each $W_\theta$ is associated with a particular decomposition of the system's dynamics.
Each decomposition, in turn, has a natural interpretation in terms of a virtual control field, ${\bm u}_\theta$.
We derive a criterion for choosing the decomposition that optimizes the convergence of the average in Eq.~\eqref{eq:ewft-intro}; and
we identify conditions under which this choice leads to a zero-variance estimator of $\Delta F$. 

Optimizing the decomposition entails inferring a time-dependent probability distribution (up to normalization) from a finite number of trajectories.
The same task arises in diffusion models~\cite{sohl2015deep,song2021scorebased,albergo2023stochastic}, a class of generative models in machine learning (ML) that provided the original inspiration for our method.
In SI-A we show that both diffusion models and our method are related to the same decomposition of overdamped Langevin dynamics. 
Hence, algorithms developed in the ML context may be leveraged to construct optimal virtual control fields.
At its core, our method involves an interplay between inference and control: information inferred from data enables the application of virtual, post-process control, without modifying the original, unescorted dynamics.

{\it Background---}Let ${\bm x}$ denote the microscopic state of a system of interest.
At temperature $T$, the energy functions $U_A({\bm x})$ and $U_B({\bm x})$ determine equilibrium states $A$ and $B$, represented by canonical distributions, $\pi_i({\bm x})=e^{[F_i-U_i({\bm x})]/T}$, $i=A,B$.
Equation~\eqref{eq:wft} applies to processes in which a system evolves from $t=0$ to $\tau$ while its energy function $U({\bm x},t)$ varies from $U(0)=U_A$ to $U(\tau)=U_B$.
Equation \eqref{eq:wft} has been derived under various dynamical schemes, including Newtonian (Hamiltonian), overdamped Langevin and underdamped Langevin dynamics~\cite{risken1989fokker, zwanzig2001nonequilibrium,gardiner1985handbook}.
Here we focus on the overdamped case.

The Langevin dynamics
\be
\label{eq:lang}
\dot{\bm x}_t = -\mu \nabla U({\bm x}_t,t) + \sqrt{2D} {\bm\xi}(t)
\ee
describe overdamped evolution in the potential $U({\bm x},t)$~\footnote{In the context of overdamped Langevin dynamics, we refer to $U({\bm x},t)$ as a ``potential'' rather than using the more general term ``energy function''.}, in the presence of a heat bath at temperature $T$.
Here, $\bm \xi$ denotes Gaussian noise satisfying $\langle \xi_i(t)\xi_j(t')\rangle = \delta_{ij}\delta(t-t')$.
The mobility and diffusion constants, $\mu$ and $D$, are related by $D=\mu T$.
The work performed on the system during the process is a functional of the trajectory, ${\bm x}_t$~\cite{Chris1997Nonequilibrium}:
\begin{equation}
\label{eq:w-unescorted}
    W[{\bm x}_t]=\int_0^\tau dt \, \partial_t U({\bm x}_t,t) \ .
\end{equation}
The angular brackets in Eq.~\eqref{eq:wft} indicate an average over an ensemble of trajectories, with initial conditions sampled from $\pi_A({\bm x})$.
We use the shorthand notation ${\bm x}_0 \sim \pi_A$ to denote this sampling.

In the escorted method, Eq.~\eqref{eq:lang} is supplemented by a user-defined, deterministic control field, ${\bm u}({\bm x},t)$:
\be
\label{eq:escorted-original}
\dot{\bm x}_t = -\mu \nabla U({\bm x}_t,t) + \sqrt{2D} \, {\bm\xi}(t) + {\bm u}({\bm x}_t,t)    \, .
\ee
The {\it escorted work} performed on the system is defined as:
\be
\label{eq:w-escorted}
W_{\bm u}[{\bm x}_t] =  \int_0^\tau dt \, \left( \partial_t U  + {\bm u} \cdot \nabla U - T \nabla\cdot{\bm u} \right) \, .
\ee
For a general choice of ${\bm u}({\bm x},t)$ (subject to mild conditions), Vaikuntanathan and Jarzynski~\cite{escortedPRL2008} derived the identity
\begin{align}
    \left\langle e^{-W_{\bm u}/T} \right\rangle_{\bm u} = e^{-\Delta F/T} \, .
\label{eq:controlFT}\end{align}
The average is taken over escorted trajectories, evolving under Eq.~\eqref{eq:escorted-original}, with ${\bm x}_0 \sim \pi_A$.

The convergence of the average in Eq.~\eqref{eq:controlFT} with the number of escorted trajectories depends on the choice of control field.
If ${\bm u}={\bm 0}$ then Eq.~\eqref{eq:controlFT} reduces to Eq.~\eqref{eq:wft}.
But if we construct ${\bm u}$ so that the system always remains in equilibrium, with respect to $U({\bm x},t)$,
then $W_{\bm u}[{\bm x}_t] = \Delta F$ for every trajectory, resulting in a zero-variance estimate of the free energy difference~\cite{escortedPRL2008}.
Thus one can accelerate convergence, relative to Eq.~\eqref{eq:wft}, by designing ${\bm u}({\bm x},t)$ so that the system remains near equilibrium throughout the process.

{\it Free energy estimation with virtual escorting---}We now show how escorting can be applied virtually -- that is, using trajectories generated without an explicit escorting field.
We continue to frame our discussion in terms of overdamped Langevin dynamics.
We extend the analysis to other dynamical schemes in SI-B. 

Suppose we have simulated, or measured, $N$ {\it unescorted} trajectories, evolving under Eq.~\eqref{eq:lang}, with ${\bm x}_0\sim\pi_A$. We wish to use these trajectories to estimate $\Delta F$.

Let $\pi_\theta({\bm x},t)$ be a strictly positive, but otherwise arbitrary, time-dependent probability distribution defined on the open time interval $(0,\tau)$.
The subscript $\theta$ denotes parameters that specify this distribution, aside from normalization.
For example, $\theta$ may denote the parameters in a neural network (NN) that outputs an unnormalized version of $\pi_\theta({\bm x},t)$.
For $t\in(0,\tau)$, we define
\begin{subequations}
\label{eq:uthetaUthetadefs}
\begin{eqnarray}
\label{eq:Utheta}
U_\theta({\bm x},t) &\equiv& -T \ln\pi_\theta({\bm x},t) + c(t) \\
\label{eq:uUUtheta}
    {\bm u}_\theta ({\bm x},t) &\equiv& -\mu\nabla\left[ U({\bm x},t)-U_\theta ({\bm x},t)\right] \, ,
\end{eqnarray}
\end{subequations}
where $c(t)$ is arbitrary and irrelevant.
We further set
\be
\label{eq:bc}
U_\theta({\bm x},0) = U_A({\bm x})
,\quad
U_\theta({\bm x},\tau) = U_B({\bm x}) \, .
\ee
$U_\theta({\bm x},t)$ may change abruptly at $t=0^+$ and $\tau^-$.

Using Eq.~\eqref{eq:uUUtheta}, we decompose Eq.~\eqref{eq:lang} as follows:
\be
\label{eq:lang-reinterpreted}
\dot{\bm x}_t = -\mu \nabla U_\theta({\bm x}_t,t) + \sqrt{2D} \, {\bm\xi}(t) + {\bm u}_\theta ({\bm x}_t,t)\, .
\ee
Here, unescorted evolution in the potential $U$ (Eq.~\eqref{eq:lang}) is reinterpreted as evolution in a virtual potential, $U_\theta$, with a virtual escorting field ${\bm u}_\theta$ -- see Eq.~\eqref{eq:escorted-original}.
We call $U_\theta$ and ${\bm u}_\theta$ ``virtual'', as they are entirely the products of our sleight-of-hand recasting of Eq.~\eqref{eq:lang} as Eq.~\eqref{eq:lang-reinterpreted}.

We now apply the escorted method to our reinterpreted dynamics.
Eqs.~\eqref{eq:w-escorted} and \eqref{eq:controlFT} become
\begin{equation}
\label{eq:Wtheta-initial}
    W_\theta[{\bm x}_t] = \int_0^\tau dt \left( \partial_t U_\theta + {\bm u}_\theta\cdot\nabla U_\theta - T \nabla\cdot {\bm u}_\theta \right)
\end{equation}
and
\be
\label{eq:expWtheta}
\left\langle e^{-W_\theta/T} \right\rangle = e^{-\Delta F/T} \, ,
\ee
where the average is over  trajectories evolving under Eq.~\eqref{eq:lang}.
Using $d_tU_\theta = \partial_t U_\theta + \dot{\bm x}_t\circ\nabla U_\theta$, where $\circ$ denotes the Stratonovich product, we rewrite Eq.~\eqref{eq:Wtheta-initial} as:
\begin{equation}
\label{eq:Wtheta}
    W_\theta[{\bm x}_t] = \Delta U + T \int_0^\tau dt \left[ \left( \dot{\bm x}_t-{\bm u}_\theta\right) \circ \nabla\ln\pi_\theta - \nabla\cdot{\bm u}_\theta \right] \, ,
\end{equation}
where $\Delta U \equiv U_B({\bm x}_\tau)-U_A({\bm x}_0)$.
With Eqs.~\eqref{eq:uthetaUthetadefs} and \eqref{eq:Wtheta}, $W_\theta$ can be evaluated, for any trajectory, provided the functions $U$ and $\nabla\ln\pi_\theta$ are known.

We arrived at Eq.~\eqref{eq:expWtheta} by appealing to Eq.~\eqref{eq:controlFT}. In SI-C we present an alternative, self-contained derivation. 
For the choice $\pi_\theta\propto e^{-U/T}$, Eq.~\eqref{eq:expWtheta} reduces to Eq.~\eqref{eq:wft}.

From Eq.~\eqref{eq:expWtheta}, we construct the estimator
\be
\Delta F_\theta^{\rm est} = -T \ln \left( \frac{1}{N} \sum_{n=1}^N e^{-W_\theta^n/T} \right) \, ,
\label{eq:estimator}\ee
where $W_\theta^n$ is the value of $W_\theta$ for the $n$'th of $N$ trajectories evolving under Eq.~\eqref{eq:lang}, with ${\bm x}_0\sim\pi_A$.
Different choices of $\pi_\theta$ produce different estimates, all of which converge to $\Delta F$ as $N\rightarrow\infty$.
We next discuss how to select $\pi_\theta$ to optimize the rate of convergence of $\Delta F_\theta^{\rm est}$.

By Jensen's inequality~\cite{jensen1906fonctions}, Eq.~\eqref{eq:expWtheta} implies non-negative {\it dissipated work}:
\be
W_\theta^{\rm diss} \equiv \langle W_\theta\rangle - \Delta F \ge 0 \, .
\ee
The convergence of exponential averages, such as those in Eqs.~\eqref{eq:wft}, \eqref{eq:controlFT} and \eqref{eq:expWtheta}, correlates strongly with dissipated work~\cite{vaikuntanathan2009dissipation}.
If $W_\theta^{\rm diss}\gg T$ then the average in Eq.~\eqref{eq:expWtheta} is dominated by unusually low values of $W_\theta$, resulting in poor convergence.
But if $W_\theta^{\rm diss}=0$ then $W_\theta = \Delta F$ for every realization of the process, resulting in a zero-variance estimator.
Hence to optimize convergence we must minimize the dissipated work.

Let $\rho({\bm x},t)$ denote the probability density that describes unescorted trajectories ${\bm x}_t$ evolving under Eq.~\eqref{eq:lang}, with ${\bm x}_0\sim\pi_A$.
This density obeys the Fokker-Planck equation,
\begin{equation}
    \label{eq:fokker-planck}
    \partial_t\rho = \mu \nabla\cdot\left(\nabla U \, \rho\right) + D\nabla^2\rho \, ,
\end{equation}
with $\rho(0)=\pi_A$.
In SI-D, we derive the result, 
\be
\label{eq:twoTerms}
W_\theta^{\rm diss} = T{\mathcal D}_{\rm KL} \left[ \rho(\tau) \, \vert\vert \, \pi_B \right] + 2DT \int_0^\tau dt \, {\mathcal D}_{\rm F} \left[ \rho(t) \, \vert\vert \, \pi_\theta(t) \right] \, .
\ee
${\mathcal D}_{\rm KL}$ and ${\mathcal D}_{\rm F}$ 
denote Kullback-Leibler and Fisher divergences~\footnote{Explicitly, ${\mathcal D}_{\rm KL}(f\vert\vert g) \equiv \int f \ln (f/g)$ and ${\mathcal D}_{\rm F}(f\vert\vert g) \equiv (1/2)\int f \vert \nabla\ln(f/g) \vert^2$. The Fisher divergence is also known as the relative Fisher information, the score-matching divergence, and the Hyv{\" a}rinen divergence.
It differs from (but is related to) the Fisher {\it information}, which involves derivatives with respect to auxiliary parameters (e.g.\ $\theta$) rather than state variables (${\bm x})$.
},
which are non-negative.
When $\pi_\theta\propto e^{-U/T}$, Eq.~\eqref{eq:twoTerms} is equivalent to Eq.~(12) of Rosa-Ra\' ices and Limmer~\cite{rosa2025nonadiabatic}.
Also, Ding, Quan and Tu~\cite{Ding2026} have recently derived fluctuation theorems related to the second term on the right in Eq.~\eqref{eq:twoTerms}.
In SI-E, we generalize Eq.~\eqref{eq:twoTerms} to the case of underdamped Langevin dynamics. 

The KL term in Eq.~\eqref{eq:twoTerms} quantifies the extent to which $\rho({\bm x},\tau)$ differs from $\pi_B({\bm x})$.
This term does not depend on the choice of $\pi_\theta$.
Since we wish to minimize $W_\theta^{\rm diss}$, the KL term represents a penalty for ending (at time $\tau$) away from equilibrium.

The Fisher term in Eq.~\eqref{eq:twoTerms} vanishes if and only if
\be
\label{eq:optimizationCondition}
\pi_\theta({\bm x},t) = \rho({\bm x},t) \quad  \textrm{for all} \,\,  t \in (0,\tau) .
\ee
We arrive at a simple conclusion: the optimal $\pi_\theta({\bm x},t)$ is just the density $\rho({\bm x},t)$ that describes an ensemble of trajectories evolving under Eq.~\eqref{eq:lang}, with ${\bm x}_0\sim\pi_A$.
What remains is to estimate this density from $N$ trajectories.

We show in SI-F that minimizing the Fisher term in Eq.~\eqref{eq:twoTerms} is equivalent to minimizing the cost function
\be
\label{eq:costFunction}
   L(\theta) = \int_0^\tau dt \int d\bm x \, \rho({\bm x},t) \left [|\nabla S_\theta (\bm x,t) |^2 -2\nabla^2 S_\theta (\bm x,t)  \right] \, ,
\ee
where $S_\theta = U_\theta/T$, i.e.\ $\pi_\theta\propto e^{-S_\theta}$.
This cost function is used in score matching, a ML strategy for training unnormalized statistical models~\cite{hyvarinen2005estimation,Song2020sliced,song2021scorebased}. 
If $\theta^\star$ denotes the parameter set that minimizes $L(\theta)$, then $\pi_{\theta^\star}$ represents the closest we can come to satisfying Eq.~\eqref{eq:optimizationCondition}.
By solving for $\theta^\star$, we optimize the convergence of $\Delta F_\theta^{\rm est}$.
In practice, we minimize $L(\theta)$ by replacing the integrals in Eq.~\eqref{eq:costFunction} by an average over trajectories.
Later, we illustrate this optimization procedure on a model system (see {\it Quartic model}). 

If $\rho({\bm x},t)$ belongs to the set of distributions $\pi_\theta({\bm x},t)$ spanned by all possible values of $\theta$, then $\pi_{\theta^\star}=\rho$. We refer to this case as {\it perfect virtual escorting}.
In this situation, the Fisher term vanishes, and $W_{\theta^\star}$ is determined uniquely by ${\bm x}_\tau$ (see SI-G): 
\begin{equation}
\label{eq:Wstar}
    W_{\theta^\star}[{\bm x}_t] = U_B({\bm x}_\tau) + T \ln \rho({\bm x}_\tau,\tau) - F_A \, .
\end{equation}
If, additionally, $\rho(\tau)=\pi_B$, then the KL term in Eq.~\eqref{eq:twoTerms} also vanishes, and $W_{\theta^\star}[{\bm x}_t] = \Delta F$ for every trajectory -- as can be seen directly from Eq.~\eqref{eq:Wstar}.
Thus, the combination of perfect virtual escorting (Eq.~\eqref{eq:optimizationCondition}) and ending in equilibrium produces a zero-variance estimator of $\Delta F$.

Even when Eq.~\eqref{eq:optimizationCondition} is imperfectly satisfied, Eq.~\eqref{eq:expWtheta} ensures that $\Delta F_\theta^{\rm est}\rightarrow\Delta F$ as $N\rightarrow\infty$.

To reduce the KL term in Eq.~\eqref{eq:twoTerms}, it may be advantageous to design $U({\bm x},t)$ to evolve in two stages: a switching stage, $0\le t\le\tau_s$, during which $U$ varies from $U_A$ to $U_B$; followed by a relaxation stage, $\tau_s < t\le\tau$, during which $U$ remains fixed at $U_B$.
During the latter, ${\mathcal D}_{\rm KL} \left[ \rho(t) \, \vert\vert \, \pi_B \right]$ decays toward zero as the system relaxes toward equilibrium.

The two-stage protocol just described highlights a difference between using Eqs.~\eqref{eq:wft} and \eqref{eq:expWtheta} to estimate $\Delta F$.
With Eq.~\eqref{eq:wft}, a relaxation stage is entirely superfluous, since no work is performed when $U({\bm x},t)$ is fixed.
Equation~\eqref{eq:expWtheta}, however, enables us to use information gathered during the relaxation stage to estimate $\Delta F$.

This difference is most evident in the limit $\tau_s \rightarrow 0$, with $\tau$ fixed.
This limit describes {\it pure relaxation} from $A$ toward $B$: from initial conditions ${\bm x}_0 \sim\pi_A$, the system simply relaxes toward $\pi_B$.
For such a process, Eq.~\eqref{eq:wft} reduces to Zwanzig's free energy perturbation formula, with $W=U_B({\bm x}_0)-U_A({\bm x}_0)$~\cite{zwanzig1954high,Chris1997Nonequilibrium}.
By contrast, from the relaxing trajectories, Eq.~\eqref{eq:expWtheta} provides a near-zero-variance estimator of $\Delta F$, if $\tau$ is sufficiently long for the system to relax close to equilibrium, and if the virtual escorting is perfect or nearly so.

\begin{figure}[t!]
	\includegraphics[width=0.48\textwidth]{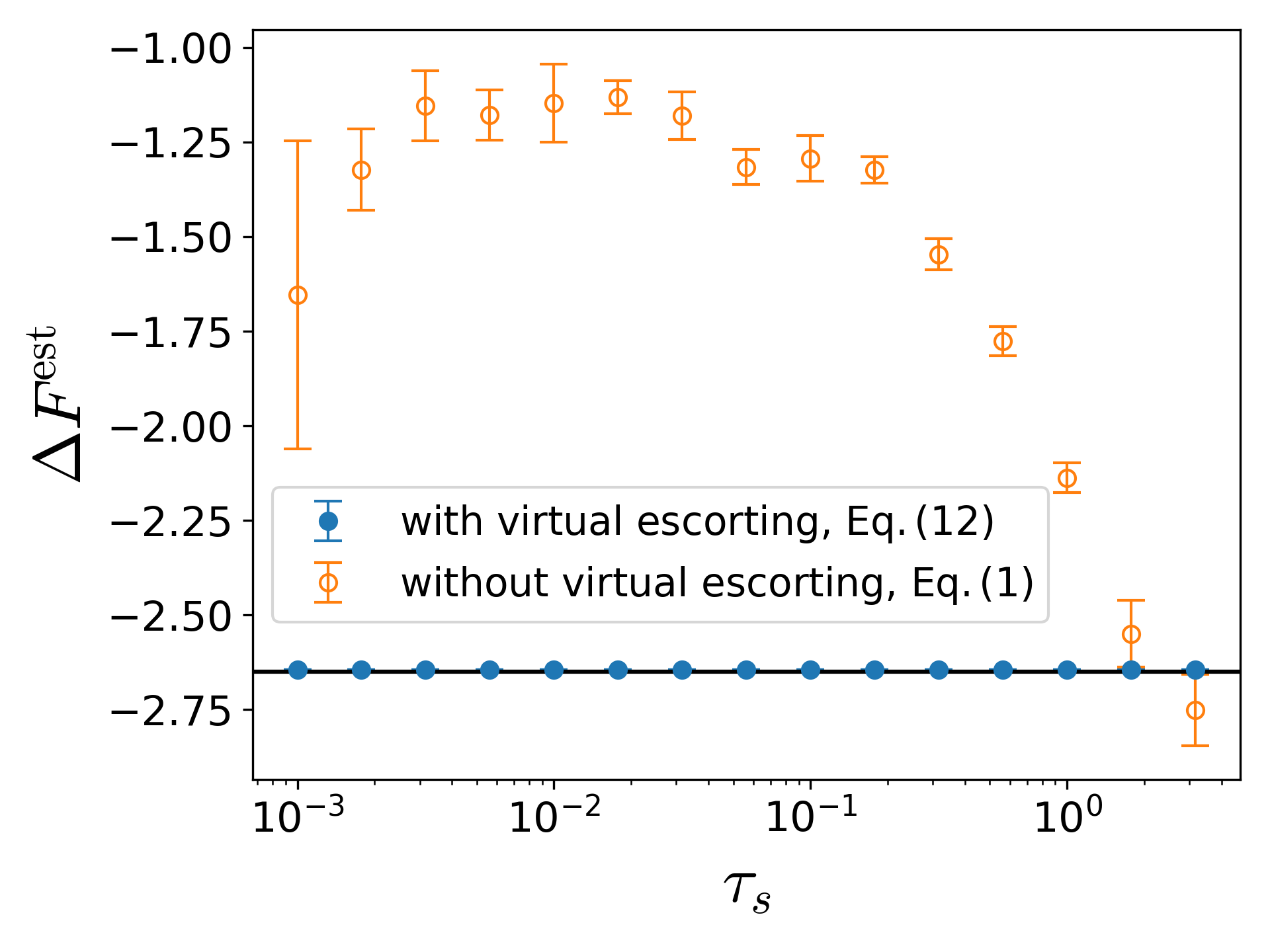}
	\vskip -0.1in
	\caption{Free energy estimation with (\textcolor{blue}{$\bullet$}) and without (\textcolor{orange}{$\circ$}) perfect virtual escorting, for a harmonic oscillator whose spring constant is switched from $k_i=100$ to $k_f=0.5$ over a switching time $\tau_s$ -- see text for details. The horizontal line indicates the true value of $\Delta F$.
	}
	\label{fig:1dHarmonic}
	\vskip -0.1in
\end{figure}

{\it Harmonic model---}
We illustrate perfect virtual escorting with an overdamped 1-D harmonic oscillator, $U=kx^2/2$.
The system evolves under the Langevin dynamics
\begin{align}
    \dot{ x}_t  = - \mu k (t) x_t + \sqrt{2\mu T } \, \xi(t) \, .
\label{eq:solvableLE}\end{align} 
The spring constant $k(t)$ satisfies
\begin{equation}
k^{-1}(t)= k_i^{-1} + (k_f^{-1} - k_i^{-1})(t/\tau_s)
\label{eq:springprotocol}
\end{equation}
during the switching stage, and $k(t)=k_f$ during the relaxation stage.
The initial and final spring constants define a free energy difference $\Delta F = 0.5 T \ln (k_f/k_i)$.

For this model, we can solve for the density $\rho(x,t)$: it is a zero-mean Gaussian with variance $\sigma^2(t)$, whose explicit form is given in SI-H. 
Setting $\pi_{\theta^\star}=\rho$, we have
$\nabla \ln \pi_{\theta^\star}=-x/\sigma^2(t)$ and $u_{\theta^\star}=\mu\left[(T/\sigma^2(t))-k\right(t)]x$.

Figure~\ref{fig:1dHarmonic} shows numerical results obtained for this model.
We simulated Eq.~\eqref{eq:solvableLE}, with $\mu,T=1$ and with initial conditions sampled from the equilibrium state at $k_i = 100$.
During each simulation, $k(t)$ was switched from $k_i$ to $k_f=0.5$ over an interval of duration $\tau_s$, as per Eq.~\eqref{eq:springprotocol}, followed by relaxation at fixed $k=k_f$ until time $\tau=5$.
For each choice of $\tau_s$, we generated $10^4$ trajectories, and estimated $\Delta F$ both without any escorting, Eq.~\eqref{eq:wft}~\footnote{
We found that the numerical evaluation of $W$ was improved by rewriting Eq.~\eqref{eq:w-unescorted} as $W[x_t] = \Delta U(x_f,x_i) - \int_0^\tau dt \, \dot x_t \circ \partial_x U(x_t,t)$, especially when $\tau_s \sim 10^{-3}$.
}, and with virtual escorting, Eq.~\eqref{eq:expWtheta}.
The trajectories were generated with a step size of $10^{-6}$, and were recorded every $100$ steps for a sampling interval of $10^{-4}$.
We evaluated $W_{\theta^\star}$ using Eq.~\eqref{eq:Wtheta}, with $\nabla \ln \pi_{\theta^\star}$ and
$u_{\theta^\star}$ as given in the previous paragraph.
We used the bootstrap method~\cite{Efron1982} to assess statistical errors.

For the parameters we chose, the system always ends close to equilibrium: ${\mathcal D}_{\rm KL} \left[ \rho(\tau) \, \vert\vert \, \pi_B \right] < 10^{-3}$ for all values of $\tau_s$ indicated in Fig.~\ref{fig:1dHarmonic}.
As a result, perfect virtual escorting yields highly accurate estimates of $\Delta F$, with bootstrapped error bars less than 0.01 \% of the estimates.
By contrast, estimates obtained using Eq.~\eqref{eq:wft} differ substantially from $\Delta F$, except at the longest switching time.
For $ \tau_s\leq 10^{-2}$, the switching stage is nearly instantaneous, which explains the plateau observed in Fig.~\ref{fig:1dHarmonic}.  Here, the system's evolution approximates a pure relaxation process, and Eq.~\eqref{eq:wft} is nearly identical to the free energy perturbation identity~\cite{zwanzig1954high}.

\begin{figure}[t]
	\includegraphics[width=0.48\textwidth]{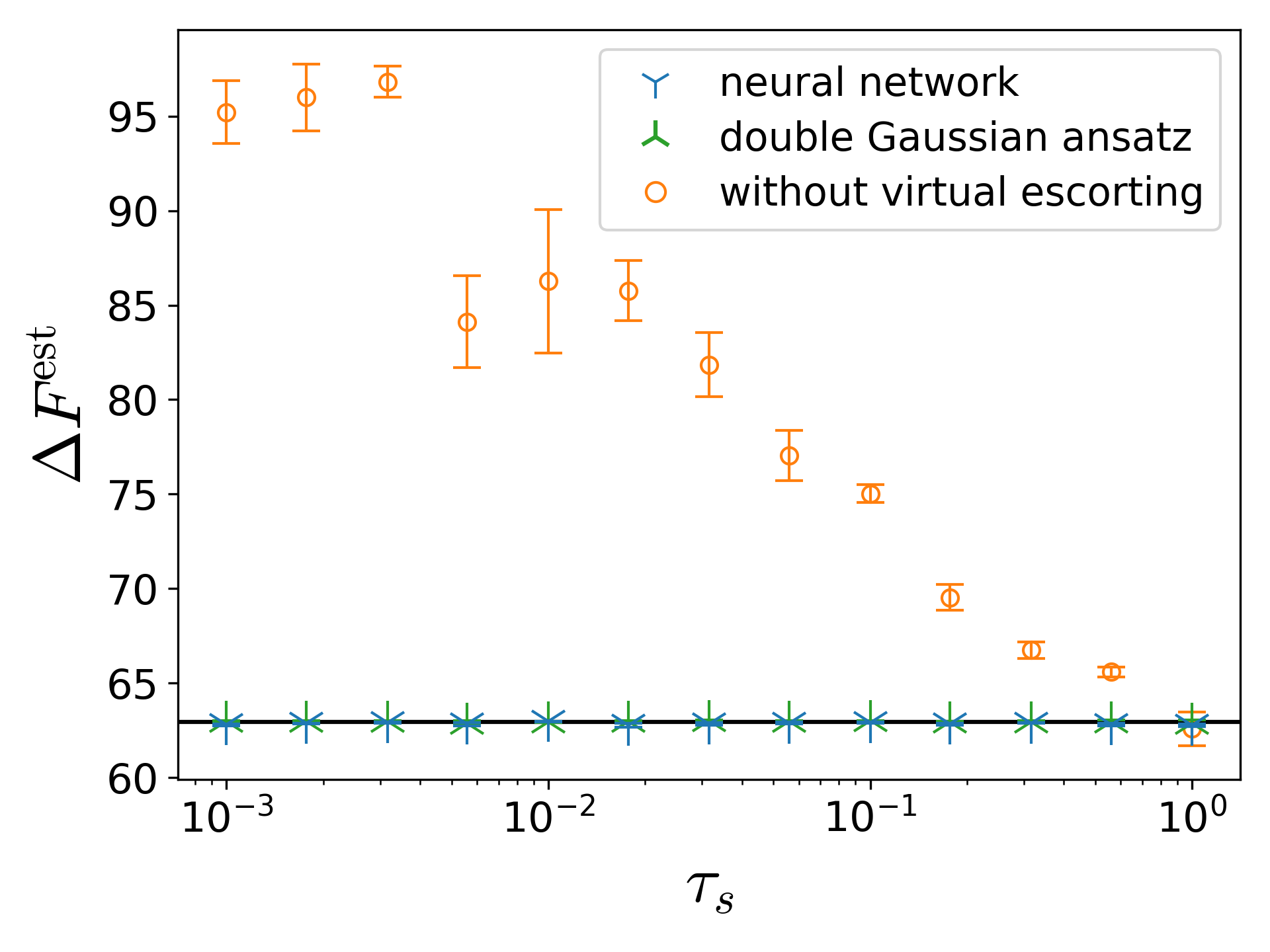}
	\vskip -0.1in
	\caption{ Free energy estimation for the quartic potential, $U(x,\lambda)=x^4-16(1-\lambda)x^2$.  The parameter $\lambda$ was varied from 0 to 1 over a time $\tau_s$, then fixed for a further interval $0.8$.
   The horizontal line indicates the known free energy difference, $\Delta F = 62.94$~\cite{oberhofer2005biased}.
   The symbols \textcolor{blue}{$\mathsf{Y}$} and \textcolor{cadmiumgreen}{$\rotatebox[origin=c]{180}{$\mathsf{Y}$}$} denote estimates obtained with Eq.~\eqref{eq:expWtheta}, using the methods for optimizing $\theta$ discussed in the text.
   At each $\tau_s$, we used the same 1000 unescorted trajectories for all three estimates (\textcolor{orange}{$\circ$}, \textcolor{blue}{$\mathsf{Y}$}, \textcolor{cadmiumgreen}{$\rotatebox[origin=c]{180}{$\mathsf{Y}$}$}), and the bootstrap method to compute statistical errors.
   }
	\label{fig:1dquartic}
	\vskip -0.1in
\end{figure}
  
{\it Quartic model---}Now consider overdamped evolution in the potential $U(x,\lambda)=x^4-16(1-\lambda)x^2$.
The parameter $\lambda$ is switched from $0$ to $1$ according to the two-stage protocol $\lambda_t={\rm min}\{(t/\tau_s), 1\}$, with $\tau_s<\tau$.
When $\lambda<1$, $U(x,\lambda)$ has minima at $\pm x_0 =\pm \sqrt{8(1-\lambda) }$, separated by a barrier of height $x_0^4$.
As $\lambda$ varies from $0$ to $1$, these minima approach one another, coalescing at the origin.
For this process, $\Delta F = 62.94 \cdots$ at $T=1$~\cite{oberhofer2005biased}.

This model process illustrates the poor convergence of Eq.~\ref{eq:wft}~\cite{oberhofer2005biased}, and the efficiency gains that can be achieved with escorting~\cite{escortedPRL2008}.
Here, we use it to illustrate virtual escorting, when $\rho(x,t)$ is not known analytically.

We simulated Eq.~\eqref{eq:lang}, with $\mu, D=1$, for values of $\tau_s$ ranging from $0.001$ to $1.0$, with $\tau=\tau_s+0.8$.
The relaxation period, of duration 0.8, is sufficient for the system to relax close to the equilibrium state $\pi_B$.
For each choice of $\tau_s$, we generated 1000 trajectories.
The trajectories were generated with a time step of $10^{-5}$, and recorded every $100$ integration steps, giving a sampling interval of $10^{-3}$.
We then tested two strategies for constructing and optimizing $\pi_\theta(x,t)$.

First, for each $\tau_s$, we used our 1000 trajectories to train a neural network (NN) to represent $S_\theta({\bm x},t)=U_\theta/T$.
The NN had four layers, with 128 units per layer, and it used the GELU activation function, $0.5 x[ 1 + {\rm erf}( {x}/{\sqrt{2}} ) ]$~\cite{hendrycks2016gaussian}.
Using the ADAM optimizer~\cite{kingma2015adam}, we minimized $L(\theta)$, Eq.~\eqref{eq:costFunction}.
(See SI-I for further details.) 
We thereby obtained parameter values $\theta^\star$ that solve, approximately, Eq.~\eqref{eq:optimizationCondition}.
We used these optimized values to compute $W_{\theta^\star}[{\bm x}_t]$ and $\Delta F_{\theta^\star}^{\rm est}$ from our simulated trajectories.

In our second strategy, we appealed to intuition and modeled $\rho(x,t)$ as a sum of two symmetrically placed Gaussian functions, reflecting the symmetry of $U(x,\lambda)$:
\begin{equation}
\label{eq:twoGaussians}
    \rho(x,t) \approx \pi_\theta(x,t) \propto e^{-(x-\bar x_t)^2/2\sigma_t^2} + e^{-(x+\bar x_t)^2/2\sigma_t^2} \, ,
\end{equation}
where $\theta = \{ \bar x_t, \sigma_t\,;\, t \in (0,\tau) \}$.
While this approximation is imperfect, determining $\theta$ is simple.
At each recorded time step, we set $\bar x_t$ and $\sigma_t$ so that the second and fourth moments of $\pi_\theta(x,t)$ matched the values $\langle x^2\rangle_t$ and $\langle x^4\rangle_t$ computed from the trajectories (see SI-J). 
We then used $\{ \bar x_t, \sigma_t \}$ to construct $W_{\theta}[{\bm x}_t]$ and $\Delta F_\theta^{\rm est}$.

Figure~\ref{fig:1dquartic} presents our results.
While Eq.~\eqref{eq:wft} converges poorly for all but the slowest simulations, Eq.~\eqref{eq:expWtheta} consistently provides accurate estimates of $\Delta F$, for both strategies discussed above.
The NN and double-Gaussian approaches produced, respectively, 0.13\,\% and 0.064\,\% average relative errors, and statistical errors less than 0.4\,\% and 0.3\,\% of the mean.
Both approaches also gave accurate estimates when using only 100 trajectories (data not shown).
Thus, from unescorted trajectories, we were able to estimate $\Delta F$ accurately with Eq.~\eqref{eq:expWtheta}, using either a NN or physical intuition (Eq.~\eqref{eq:twoGaussians}) to model $\rho(x,t)$.

{\it Discussion--}We have introduced a free-energy estimation method that draws on the escorted approach of Ref.~\cite{escortedPRL2008}, without requiring trajectories that actually evolve under a control field, ${\bm u}$.
Instead, our method uses unescorted trajectories, and a {\it virtual} control field constructed from a model $e^{-S_\theta({\bm x},t)} \propto \pi_\theta({\bm x},t)$.
A feature of our method is that convergence is optimized {\it after} the trajectories have been generated.

Optimal convergence of Eq.~\eqref{eq:expWtheta} occurs when $\pi_\theta({\bm x},t)$ accurately describes the evolving ensemble of trajectories, Eq.~\eqref{eq:optimizationCondition}.
If the system ends near equilibrium, then the optimal parameters $\theta^\star$ produce a near-zero-variance estimate of $\Delta F$.
This feature is illustrated in Figs.~\ref{fig:1dHarmonic} and \ref{fig:1dquartic}, where our method produces highly accurate estimates of $\Delta F$, using trajectories for which Eq.~\eqref{eq:wft} converges poorly.

With our method, $\Delta F$ can be estimated using trajectories generated during a pure relaxation process, in which the system relaxes from state $A$ toward state $B$.
If $U_B({\bm x})$ is quadratic, these trajectories evolve under the Ornstein-Uhlenbeck (OU) process, whose propagator is known~\cite{gardiner1985handbook,risken1989fokker}.
In this context, our method might offer a twist on the widely used Frenkel-Ladd method of estimating free energies of crystalline solids, in which $U_A$ describes the solid in question, and $U_B$ its Einstein crystal counterpart~\cite{Frenkel1984}.
With our method, $\Delta F$ can be estimated from trajectories that simply relax from $A$ to $B$ under the OU process.

By Eqs.~\eqref{eq:uthetaUthetadefs} and \eqref{eq:Wtheta}, the flow field ${\bm u}_\theta({\bm x},t)$ uniquely specifies the functional $W_\theta[{\bm x}_t]$.
Thus, we could implement our method by modeling and optimizing ${\bm u}_\theta$, rather than $\pi_\theta$.
In generative ML, this approach is known as flow matching~\cite{Lipman2023,Klinger2025,Ikeda2025}.
The optimality condition on ${\bm u}_\theta$ is (see SI-K): 
\begin{equation}
    \label{eq:continuity}
    \partial_t\rho = -\nabla\cdot({\bm u}_{\theta^\star}\rho) \, .
\end{equation}
Hence the flow $\dot{\bm x}_t = {\bm u}_{\theta^\star}({\bm x}_t,t)$ generates the same density, $\rho({\bm x},t)$, as the Langevin dynamics of Eq.~\eqref{eq:lang}, when ${\bm x}_0\sim \pi_A$.

In Eqs.~\eqref{eq:Wtheta} and \eqref{eq:costFunction}, the computation of $W_\theta$ and $L(\theta)$ involves evaluating the Laplacian $\nabla^2S_\theta$~\footnote{
In Eq.~\eqref{eq:Wtheta}, $\nabla\cdot{\bm u}_\theta=\mu T \nabla^2 S_\theta - \mu\nabla^2 U$.
}, which can be costly when $S_\theta$ is represented by a NN.
In SI-L we show that this cost can be avoided: both $W_\theta$ and $L(\theta)$ can be computed from evaluations of $\nabla S_\theta$ rather than $\nabla^2 S_\theta$. 

Finally, it will be interesting to investigate whether our method can be applied to experimental data, if we have access only to one or a few degrees of freedom, whose evolution may accurately be modeled by overdamped Langevin dynamics.

\begin{acknowledgments}
{\it Acknowledgments---}This research was supported by the KIAS Individual Grant Nos. PG081802 (S.L.) and the Center for Advanced Computation at the Korea Institute for Advanced Study.
S.L. thanks L. H. Cheung for discussions and verification in the early stage of this research, Y.~K. Noh for comments on training a neural network and M. te Vrugt for discussion on possible applications.
C.J. thanks D.\ Limmer, J.\ Sohl-Dickstein, M.\ Welling, P. Tiwary, and members of the Tiwary research group, for useful comments and conversations.
\end{acknowledgments}

\bibliography{Diffusion}

\begin{thebibliography}{50}%
\makeatletter
\providecommand \@ifxundefined [1]{%
 \@ifx{#1\undefined}
}%
\providecommand \@ifnum [1]{%
 \ifnum #1\expandafter \@firstoftwo
 \else \expandafter \@secondoftwo
 \fi
}%
\providecommand \@ifx [1]{%
 \ifx #1\expandafter \@firstoftwo
 \else \expandafter \@secondoftwo
 \fi
}%
\providecommand \natexlab [1]{#1}%
\providecommand \enquote  [1]{``#1''}%
\providecommand \bibnamefont  [1]{#1}%
\providecommand \bibfnamefont [1]{#1}%
\providecommand \citenamefont [1]{#1}%
\providecommand \href@noop [0]{\@secondoftwo}%
\providecommand \href [0]{\begingroup \@sanitize@url \@href}%
\providecommand \@href[1]{\@@startlink{#1}\@@href}%
\providecommand \@@href[1]{\endgroup#1\@@endlink}%
\providecommand \@sanitize@url [0]{\catcode `\\12\catcode `\$12\catcode
  `\&12\catcode `\#12\catcode `\^12\catcode `\_12\catcode `\%12\relax}%
\providecommand \@@startlink[1]{}%
\providecommand \@@endlink[0]{}%
\providecommand \url  [0]{\begingroup\@sanitize@url \@url }%
\providecommand \@url [1]{\endgroup\@href {#1}{\urlprefix }}%
\providecommand \urlprefix  [0]{URL }%
\providecommand \Eprint [0]{\href }%
\providecommand \doibase [0]{http://dx.doi.org/}%
\providecommand \selectlanguage [0]{\@gobble}%
\providecommand \bibinfo  [0]{\@secondoftwo}%
\providecommand \bibfield  [0]{\@secondoftwo}%
\providecommand \translation [1]{[#1]}%
\providecommand \BibitemOpen [0]{}%
\providecommand \bibitemStop [0]{}%
\providecommand \bibitemNoStop [0]{.\EOS\space}%
\providecommand \EOS [0]{\spacefactor3000\relax}%
\providecommand \BibitemShut  [1]{\csname bibitem#1\endcsname}%
\let\auto@bib@innerbib\@empty
\bibitem [{\citenamefont
  {Jarzynski}(1997{\natexlab{a}})}]{Chris1997Nonequilibrium}%
  \BibitemOpen
  \bibfield  {author} {\bibinfo {author} {\bibfnamefont {C.}~\bibnamefont
  {Jarzynski}},\ }\href {\doibase 10.1103/PhysRevLett.78.2690} {\bibfield
  {journal} {\bibinfo  {journal} {Phys. Rev. Lett.}\ }\textbf {\bibinfo
  {volume} {78}},\ \bibinfo {pages} {2690} (\bibinfo {year}
  {1997}{\natexlab{a}})}\BibitemShut {NoStop}%
\bibitem [{\citenamefont {Jarzynski}(1997{\natexlab{b}})}]{ChrisPRE1997}%
  \BibitemOpen
  \bibfield  {author} {\bibinfo {author} {\bibfnamefont {C.}~\bibnamefont
  {Jarzynski}},\ }\href {\doibase 10.1103/PhysRevE.56.5018} {\bibfield
  {journal} {\bibinfo  {journal} {Phys. Rev. E}\ }\textbf {\bibinfo {volume}
  {56}},\ \bibinfo {pages} {5018} (\bibinfo {year}
  {1997}{\natexlab{b}})}\BibitemShut {NoStop}%
\bibitem [{\citenamefont {Vaikuntanathan}\ and\ \citenamefont
  {Jarzynski}(2009)}]{vaikuntanathan2009dissipation}%
  \BibitemOpen
  \bibfield  {author} {\bibinfo {author} {\bibfnamefont {S.}~\bibnamefont
  {Vaikuntanathan}}\ and\ \bibinfo {author} {\bibfnamefont {C.}~\bibnamefont
  {Jarzynski}},\ }\href@noop {} {\bibfield  {journal} {\bibinfo  {journal}
  {Europhys. Lett.}\ }\textbf {\bibinfo {volume} {87}},\ \bibinfo {pages}
  {60005} (\bibinfo {year} {2009})}\BibitemShut {NoStop}%
\bibitem [{\citenamefont {Vaikuntanathan}\ and\ \citenamefont
  {Jarzynski}(2008)}]{escortedPRL2008}%
  \BibitemOpen
  \bibfield  {author} {\bibinfo {author} {\bibfnamefont {S.}~\bibnamefont
  {Vaikuntanathan}}\ and\ \bibinfo {author} {\bibfnamefont {C.}~\bibnamefont
  {Jarzynski}},\ }\href {\doibase 10.1103/PhysRevLett.100.190601} {\bibfield
  {journal} {\bibinfo  {journal} {Phys. Rev. Lett.}\ }\textbf {\bibinfo
  {volume} {100}},\ \bibinfo {pages} {190601} (\bibinfo {year}
  {2008})}\BibitemShut {NoStop}%
\bibitem [{\citenamefont {Wirnsberger}\ \emph {et~al.}(2020)\citenamefont
  {Wirnsberger}, \citenamefont {Ballard}, \citenamefont {Papamakarios},
  \citenamefont {Abercrombie}, \citenamefont {Racani{\`e}re}, \citenamefont
  {Pritzel}, \citenamefont {Jimenez~Rezende},\ and\ \citenamefont
  {Blundell}}]{Wirnsberger2020TargetedML}%
  \BibitemOpen
  \bibfield  {author} {\bibinfo {author} {\bibfnamefont {P.}~\bibnamefont
  {Wirnsberger}}, \bibinfo {author} {\bibfnamefont {A.~J.}\ \bibnamefont
  {Ballard}}, \bibinfo {author} {\bibfnamefont {G.}~\bibnamefont
  {Papamakarios}}, \bibinfo {author} {\bibfnamefont {S.}~\bibnamefont
  {Abercrombie}}, \bibinfo {author} {\bibfnamefont {S.}~\bibnamefont
  {Racani{\`e}re}}, \bibinfo {author} {\bibfnamefont {A.}~\bibnamefont
  {Pritzel}}, \bibinfo {author} {\bibfnamefont {D.}~\bibnamefont
  {Jimenez~Rezende}}, \ and\ \bibinfo {author} {\bibfnamefont {C.}~\bibnamefont
  {Blundell}},\ }\href {\doibase 10.1063/5.0018903} {\bibfield  {journal}
  {\bibinfo  {journal} {J. Chem. Phys.}\ }\textbf {\bibinfo {volume} {153}},\
  \bibinfo {pages} {144112} (\bibinfo {year} {2020})}\BibitemShut {NoStop}%
\bibitem [{\citenamefont {Vargas}\ \emph {et~al.}(2024)\citenamefont {Vargas},
  \citenamefont {Padhy}, \citenamefont {Blessing},\ and\ \citenamefont
  {N{\"u}sken}}]{vargas2024transport}%
  \BibitemOpen
  \bibfield  {author} {\bibinfo {author} {\bibfnamefont {F.}~\bibnamefont
  {Vargas}}, \bibinfo {author} {\bibfnamefont {S.}~\bibnamefont {Padhy}},
  \bibinfo {author} {\bibfnamefont {D.}~\bibnamefont {Blessing}}, \ and\
  \bibinfo {author} {\bibfnamefont {N.}~\bibnamefont {N{\"u}sken}},\ }in\ \href
  {https://openreview.net/forum?id=PP1rudnxiW} {\emph {\bibinfo {booktitle}
  {International Conference on Learning Representations (ICLR)}}}\ (\bibinfo
  {year} {2024})\BibitemShut {NoStop}%
\bibitem [{\citenamefont {Zhong}\ \emph {et~al.}(2024)\citenamefont {Zhong},
  \citenamefont {Kuznets-Speck},\ and\ \citenamefont
  {DeWeese}}]{Zhong2024TimeAsymmetric}%
  \BibitemOpen
  \bibfield  {author} {\bibinfo {author} {\bibfnamefont {A.}~\bibnamefont
  {Zhong}}, \bibinfo {author} {\bibfnamefont {B.}~\bibnamefont
  {Kuznets-Speck}}, \ and\ \bibinfo {author} {\bibfnamefont {M.~R.}\
  \bibnamefont {DeWeese}},\ }\href {\doibase 10.1103/PhysRevE.110.034121}
  {\bibfield  {journal} {\bibinfo  {journal} {Phys. Rev. E}\ }\textbf {\bibinfo
  {volume} {110}},\ \bibinfo {pages} {034121} (\bibinfo {year}
  {2024})}\BibitemShut {NoStop}%
\bibitem [{\citenamefont {M{\'a}t{\'e}}\ \emph {et~al.}(2024)\citenamefont
  {M{\'a}t{\'e}}, \citenamefont {Fleuret},\ and\ \citenamefont
  {Bereau}}]{Mate2024Neural}%
  \BibitemOpen
  \bibfield  {author} {\bibinfo {author} {\bibfnamefont {B.}~\bibnamefont
  {M{\'a}t{\'e}}}, \bibinfo {author} {\bibfnamefont {F.}~\bibnamefont
  {Fleuret}}, \ and\ \bibinfo {author} {\bibfnamefont {T.}~\bibnamefont
  {Bereau}},\ }\href {\doibase 10.1021/acs.jpclett.4c01958} {\bibfield
  {journal} {\bibinfo  {journal} {J. Phys. Chem. Lett.}\ }\textbf {\bibinfo
  {volume} {15}},\ \bibinfo {pages} {11395} (\bibinfo {year}
  {2024})}\BibitemShut {NoStop}%
\bibitem [{\citenamefont {M{\'a}t{\'e}}\ \emph {et~al.}(2025)\citenamefont
  {M{\'a}t{\'e}}, \citenamefont {Fleuret},\ and\ \citenamefont
  {Bereau}}]{Mate2025}%
  \BibitemOpen
  \bibfield  {author} {\bibinfo {author} {\bibfnamefont {B.}~\bibnamefont
  {M{\'a}t{\'e}}}, \bibinfo {author} {\bibfnamefont {F.}~\bibnamefont
  {Fleuret}}, \ and\ \bibinfo {author} {\bibfnamefont {T.}~\bibnamefont
  {Bereau}},\ }\href@noop {} {\bibfield  {journal} {\bibinfo  {journal} {J.
  Chem. Phys.}\ }\textbf {\bibinfo {volume} {162}},\ \bibinfo {pages} {124107}
  (\bibinfo {year} {2025})}\BibitemShut {NoStop}%
\bibitem [{\citenamefont {Rosa-Ra{\'\i}ces}\ and\ \citenamefont
  {Limmer}(2025)}]{rosa2025nonadiabatic}%
  \BibitemOpen
  \bibfield  {author} {\bibinfo {author} {\bibfnamefont {J.~L.}\ \bibnamefont
  {Rosa-Ra{\'\i}ces}}\ and\ \bibinfo {author} {\bibfnamefont {D.~T.}\
  \bibnamefont {Limmer}},\ }\href@noop {} {\bibfield  {journal} {\bibinfo
  {journal} {J. Chem. Theory Comput.}\ }\textbf {\bibinfo {volume} {21}},\
  \bibinfo {pages} {11455} (\bibinfo {year} {2025})}\BibitemShut {NoStop}%
\bibitem [{\citenamefont {He}\ \emph {et~al.}(2025)\citenamefont {He},
  \citenamefont {Du}, \citenamefont {Vargas}, \citenamefont {Wang},
  \citenamefont {Gomes}, \citenamefont {Hern{\'a}ndez-Lobato},\ and\
  \citenamefont {Vanden-Eijnden}}]{he2025feat}%
  \BibitemOpen
  \bibfield  {author} {\bibinfo {author} {\bibfnamefont {J.}~\bibnamefont
  {He}}, \bibinfo {author} {\bibfnamefont {Y.}~\bibnamefont {Du}}, \bibinfo
  {author} {\bibfnamefont {F.}~\bibnamefont {Vargas}}, \bibinfo {author}
  {\bibfnamefont {Y.}~\bibnamefont {Wang}}, \bibinfo {author} {\bibfnamefont
  {C.~P.}\ \bibnamefont {Gomes}}, \bibinfo {author} {\bibfnamefont {J.~M.}\
  \bibnamefont {Hern{\'a}ndez-Lobato}}, \ and\ \bibinfo {author} {\bibfnamefont
  {E.}~\bibnamefont {Vanden-Eijnden}},\ }in\ \href@noop {} {\emph {\bibinfo
  {booktitle} {Advances in Neural Information Processing Systems (NeurIPS)}}}\
  (\bibinfo {year} {2025})\BibitemShut {NoStop}%
\bibitem [{\citenamefont {Holdijk}\ \emph {et~al.}(2026)\citenamefont
  {Holdijk}, \citenamefont {Anand}, \citenamefont {Bronstein},\ and\
  \citenamefont {Welling}}]{Holdijk2026}%
  \BibitemOpen
  \bibfield  {author} {\bibinfo {author} {\bibfnamefont {L.}~\bibnamefont
  {Holdijk}}, \bibinfo {author} {\bibfnamefont {N.~M.}\ \bibnamefont {Anand}},
  \bibinfo {author} {\bibfnamefont {M.~M.}\ \bibnamefont {Bronstein}}, \ and\
  \bibinfo {author} {\bibfnamefont {M.}~\bibnamefont {Welling}},\ }in\
  \href@noop {} {\emph {\bibinfo {booktitle} {International Conference on
  Learning Representations (ICLR)}}}\ (\bibinfo {year} {2026})\BibitemShut
  {NoStop}%
\bibitem [{\citenamefont {Schebek}\ \emph {et~al.}(2026)\citenamefont
  {Schebek}, \citenamefont {He}, \citenamefont {Hoffmann}, \citenamefont {Du},
  \citenamefont {No{\' e}},\ and\ \citenamefont {Rogal}}]{Schebek2026}%
  \BibitemOpen
  \bibfield  {author} {\bibinfo {author} {\bibfnamefont {M.}~\bibnamefont
  {Schebek}}, \bibinfo {author} {\bibfnamefont {J.}~\bibnamefont {He}},
  \bibinfo {author} {\bibfnamefont {E.}~\bibnamefont {Hoffmann}}, \bibinfo
  {author} {\bibfnamefont {Y.}~\bibnamefont {Du}}, \bibinfo {author}
  {\bibfnamefont {F.}~\bibnamefont {No{\' e}}}, \ and\ \bibinfo {author}
  {\bibfnamefont {J.}~\bibnamefont {Rogal}},\ }\href@noop {} {\bibfield
  {journal} {\bibinfo  {journal} {J. Chem. Phys.}\ }\textbf {\bibinfo {volume}
  {164}},\ \bibinfo {pages} {184111} (\bibinfo {year} {2026})}\BibitemShut
  {NoStop}%
\bibitem [{\citenamefont {Sohl-Dickstein}\ \emph {et~al.}(2015)\citenamefont
  {Sohl-Dickstein}, \citenamefont {Weiss}, \citenamefont {Maheswaranathan},\
  and\ \citenamefont {Ganguli}}]{sohl2015deep}%
  \BibitemOpen
  \bibfield  {author} {\bibinfo {author} {\bibfnamefont {J.}~\bibnamefont
  {Sohl-Dickstein}}, \bibinfo {author} {\bibfnamefont {E.~A.}\ \bibnamefont
  {Weiss}}, \bibinfo {author} {\bibfnamefont {N.}~\bibnamefont
  {Maheswaranathan}}, \ and\ \bibinfo {author} {\bibfnamefont {S.}~\bibnamefont
  {Ganguli}},\ }in\ \href@noop {} {\emph {\bibinfo {booktitle} {International
  Conference on Machine Learning (ICML)}}}\ (\bibinfo {year}
  {2015})\BibitemShut {NoStop}%
\bibitem [{\citenamefont {Song}\ \emph {et~al.}(2021)\citenamefont {Song},
  \citenamefont {Sohl-Dickstein}, \citenamefont {Kingma}, \citenamefont
  {Kumar}, \citenamefont {Ermon},\ and\ \citenamefont
  {Poole}}]{song2021scorebased}%
  \BibitemOpen
  \bibfield  {author} {\bibinfo {author} {\bibfnamefont {Y.}~\bibnamefont
  {Song}}, \bibinfo {author} {\bibfnamefont {J.}~\bibnamefont
  {Sohl-Dickstein}}, \bibinfo {author} {\bibfnamefont {D.~P.}\ \bibnamefont
  {Kingma}}, \bibinfo {author} {\bibfnamefont {A.}~\bibnamefont {Kumar}},
  \bibinfo {author} {\bibfnamefont {S.}~\bibnamefont {Ermon}}, \ and\ \bibinfo
  {author} {\bibfnamefont {B.}~\bibnamefont {Poole}},\ }in\ \href@noop {}
  {\emph {\bibinfo {booktitle} {Advances in Neural Information Processing
  Systems (NeurIPS)}}}\ (\bibinfo {year} {2021})\BibitemShut {NoStop}%
\bibitem [{\citenamefont {Albergo}\ \emph {et~al.}(2025)\citenamefont
  {Albergo}, \citenamefont {Boffi},\ and\ \citenamefont
  {Vanden-Eijnden}}]{albergo2023stochastic}%
  \BibitemOpen
  \bibfield  {author} {\bibinfo {author} {\bibfnamefont {M.}~\bibnamefont
  {Albergo}}, \bibinfo {author} {\bibfnamefont {N.~M.}\ \bibnamefont {Boffi}},
  \ and\ \bibinfo {author} {\bibfnamefont {E.}~\bibnamefont {Vanden-Eijnden}},\
  }\href@noop {} {\bibfield  {journal} {\bibinfo  {journal} {J. Mach. Learn.
  Technol.}\ }\textbf {\bibinfo {volume} {26}},\ \bibinfo {pages} {1} (\bibinfo
  {year} {2025})}\BibitemShut {NoStop}%
\bibitem [{\citenamefont {Risken}(1989)}]{risken1989fokker}%
  \BibitemOpen
  \bibfield  {author} {\bibinfo {author} {\bibfnamefont {H.}~\bibnamefont
  {Risken}},\ }in\ \href@noop {} {\emph {\bibinfo {booktitle} {The
  Fokker-Planck equation: methods of solution and applications}}}\ (\bibinfo
  {publisher} {Springer},\ \bibinfo {year} {1989})\BibitemShut {NoStop}%
\bibitem [{\citenamefont {Zwanzig}(2001)}]{zwanzig2001nonequilibrium}%
  \BibitemOpen
  \bibfield  {author} {\bibinfo {author} {\bibfnamefont {R.}~\bibnamefont
  {Zwanzig}},\ }\href@noop {} {\emph {\bibinfo {title} {Nonequilibrium
  Statistical Mechanics}}}\ (\bibinfo  {publisher} {Oxford University Press},\
  \bibinfo {year} {2001})\BibitemShut {NoStop}%
\bibitem [{\citenamefont {Gardiner}\ \emph {et~al.}(1985)\citenamefont
  {Gardiner} \emph {et~al.}}]{gardiner1985handbook}%
  \BibitemOpen
  \bibfield  {author} {\bibinfo {author} {\bibfnamefont {C.~W.}\ \bibnamefont
  {Gardiner}} \emph {et~al.},\ }\href@noop {} {\emph {\bibinfo {title}
  {Handbook of Stochastic Methods}}}\ (\bibinfo  {publisher} {Springer},\
  \bibinfo {year} {1985})\BibitemShut {NoStop}%
\bibitem [{Note1()}]{Note1}%
  \BibitemOpen
  \bibinfo {note} {In the context of overdamped Langevin dynamics, we refer to
  $U({\protect \bm {x}},t)$ as a ``potential'' rather than using the more
  general term ``energy function''.}\BibitemShut {Stop}%
\bibitem [{\citenamefont {Jensen}(1906)}]{jensen1906fonctions}%
  \BibitemOpen
  \bibfield  {author} {\bibinfo {author} {\bibfnamefont {J.~L. W.~V.}\
  \bibnamefont {Jensen}},\ }\href@noop {} {\bibfield  {journal} {\bibinfo
  {journal} {Acta Mathematica}\ }\textbf {\bibinfo {volume} {30}},\ \bibinfo
  {pages} {175} (\bibinfo {year} {1906})}\BibitemShut {NoStop}%
\bibitem [{Note2()}]{Note2}%
  \BibitemOpen
  \bibinfo {note} {Explicitly, ${\protect \mathcal D}_{\protect \rm KL}(f\vert
  \vert g) \equiv \DOTSI \intop \ilimits@ f \ln (f/g)$ and ${\protect \mathcal
  D}_{\protect \rm F}(f\vert \vert g) \equiv (1/2)\DOTSI \intop \ilimits@ f
  \vert \nabla \ln (f/g) \vert ^2$. The Fisher divergence is also known as the
  relative Fisher information, the score-matching divergence, and the Hyv{\"
  a}rinen divergence. It differs from (but is related to) the Fisher {\protect
  \it information}, which involves derivatives with respect to auxiliary
  parameters (e.g.\ $\theta $) rather than state variables (${\protect \bm
  {x}})$.}\BibitemShut {Stop}%
\bibitem [{\citenamefont {Ding}\ \emph {et~al.}(2026)\citenamefont {Ding},
  \citenamefont {Quan},\ and\ \citenamefont {Tu}}]{Ding2026}%
  \BibitemOpen
  \bibfield  {author} {\bibinfo {author} {\bibfnamefont {X.}~\bibnamefont
  {Ding}}, \bibinfo {author} {\bibfnamefont {H.~T.}\ \bibnamefont {Quan}}, \
  and\ \bibinfo {author} {\bibfnamefont {Y.}~\bibnamefont {Tu}},\ }\href@noop
  {} {\  (\bibinfo {year} {2026})},\ \Eprint {http://arxiv.org/abs/2606.17252}
  {arXiv:2606.17252} \BibitemShut {NoStop}%
\bibitem [{\citenamefont {Hyv{\"a}rinen}\ and\ \citenamefont
  {Dayan}(2005)}]{hyvarinen2005estimation}%
  \BibitemOpen
  \bibfield  {author} {\bibinfo {author} {\bibfnamefont {A.}~\bibnamefont
  {Hyv{\"a}rinen}}\ and\ \bibinfo {author} {\bibfnamefont {P.}~\bibnamefont
  {Dayan}},\ }\href@noop {} {\bibfield  {journal} {\bibinfo  {journal} {J.
  Mach. Learn. Res.}\ }\textbf {\bibinfo {volume} {6}} (\bibinfo {year}
  {2005})}\BibitemShut {NoStop}%
\bibitem [{\citenamefont {Song}\ \emph {et~al.}(2020)\citenamefont {Song},
  \citenamefont {Garg}, \citenamefont {Shi},\ and\ \citenamefont
  {Ermon}}]{Song2020sliced}%
  \BibitemOpen
  \bibfield  {author} {\bibinfo {author} {\bibfnamefont {Y.}~\bibnamefont
  {Song}}, \bibinfo {author} {\bibfnamefont {S.}~\bibnamefont {Garg}}, \bibinfo
  {author} {\bibfnamefont {J.}~\bibnamefont {Shi}}, \ and\ \bibinfo {author}
  {\bibfnamefont {S.}~\bibnamefont {Ermon}},\ }in\ \href@noop {} {\emph
  {\bibinfo {booktitle} {Uncertainty in Artificial Intelligence Conference
  (UAI)}}}\ (\bibinfo {year} {2020})\BibitemShut {NoStop}%
\bibitem [{\citenamefont {Zwanzig}(1954)}]{zwanzig1954high}%
  \BibitemOpen
  \bibfield  {author} {\bibinfo {author} {\bibfnamefont {R.~W.}\ \bibnamefont
  {Zwanzig}},\ }\href@noop {} {\bibfield  {journal} {\bibinfo  {journal} {J.
  Chem. Phys.}\ }\textbf {\bibinfo {volume} {22}},\ \bibinfo {pages} {1420}
  (\bibinfo {year} {1954})}\BibitemShut {NoStop}%
\bibitem [{Note3()}]{Note3}%
  \BibitemOpen
  \bibinfo {note} {We found that the numerical evaluation of $W$ was improved
  by rewriting Eq.~\protect \eqref {eq:w-unescorted} as $W[x_t] = \Delta
  U(x_f,x_i) - \DOTSI \intop \ilimits@ _0^\tau dt \protect \, \protect \dot x_t
  \circ \partial _x U(x_t,t)$, especially when $\tau _s \sim
  10^{-3}$.}\BibitemShut {Stop}%
\bibitem [{\citenamefont {Efron}(1982)}]{Efron1982}%
  \BibitemOpen
  \bibfield  {author} {\bibinfo {author} {\bibfnamefont {B.}~\bibnamefont
  {Efron}},\ }\href@noop {} {\emph {\bibinfo {title} {The Jackknife, the
  Bootstrap, and Other Resampling Plans}}}\ (\bibinfo  {publisher} {Society for
  Industrial and Applied Mathematics (SIAM)},\ \bibinfo {year}
  {1982})\BibitemShut {NoStop}%
\bibitem [{\citenamefont {Oberhofer}\ \emph {et~al.}(2005)\citenamefont
  {Oberhofer}, \citenamefont {Dellago},\ and\ \citenamefont
  {Geissler}}]{oberhofer2005biased}%
  \BibitemOpen
  \bibfield  {author} {\bibinfo {author} {\bibfnamefont {H.}~\bibnamefont
  {Oberhofer}}, \bibinfo {author} {\bibfnamefont {C.}~\bibnamefont {Dellago}},
  \ and\ \bibinfo {author} {\bibfnamefont {P.~L.}\ \bibnamefont {Geissler}},\
  }\href@noop {} {\bibfield  {journal} {\bibinfo  {journal} {J. Phys. Chem. B}\
  }\textbf {\bibinfo {volume} {109}},\ \bibinfo {pages} {6902} (\bibinfo {year}
  {2005})}\BibitemShut {NoStop}%
\bibitem [{\citenamefont {Hendrycks}\ and\ \citenamefont
  {Gimpel}(2016)}]{hendrycks2016gaussian}%
  \BibitemOpen
  \bibfield  {author} {\bibinfo {author} {\bibfnamefont {D.}~\bibnamefont
  {Hendrycks}}\ and\ \bibinfo {author} {\bibfnamefont {K.}~\bibnamefont
  {Gimpel}},\ }\href@noop {} {\  (\bibinfo {year} {2016})},\ \Eprint
  {http://arxiv.org/abs/1606.08415} {arXiv:1606.08415} \BibitemShut {NoStop}%
\bibitem [{\citenamefont {Kingma}\ and\ \citenamefont
  {Ba}(2015)}]{kingma2015adam}%
  \BibitemOpen
  \bibfield  {author} {\bibinfo {author} {\bibfnamefont {D.~P.}\ \bibnamefont
  {Kingma}}\ and\ \bibinfo {author} {\bibfnamefont {J.}~\bibnamefont {Ba}},\
  }in\ \href@noop {} {\emph {\bibinfo {booktitle} {International Conference on
  Learning Representations (ICLR)}}}\ (\bibinfo {year} {2015})\BibitemShut
  {NoStop}%
\bibitem [{\citenamefont {Frenkel}\ and\ \citenamefont
  {Ladd}(1984)}]{Frenkel1984}%
  \BibitemOpen
  \bibfield  {author} {\bibinfo {author} {\bibfnamefont {D.}~\bibnamefont
  {Frenkel}}\ and\ \bibinfo {author} {\bibfnamefont {A.~J.}\ \bibnamefont
  {Ladd}},\ }\href@noop {} {\bibfield  {journal} {\bibinfo  {journal} {J. Chem.
  Phys}\ }\textbf {\bibinfo {volume} {81}},\ \bibinfo {pages} {3188 } (\bibinfo
  {year} {1984})}\BibitemShut {NoStop}%
\bibitem [{\citenamefont {Lipman}\ \emph {et~al.}(2023)\citenamefont {Lipman},
  \citenamefont {Chen}, \citenamefont {Ben-Hamu}, \citenamefont {Nickel},\ and\
  \citenamefont {Le}}]{Lipman2023}%
  \BibitemOpen
  \bibfield  {author} {\bibinfo {author} {\bibfnamefont {Y.}~\bibnamefont
  {Lipman}}, \bibinfo {author} {\bibfnamefont {R.~T.~Q.}\ \bibnamefont {Chen}},
  \bibinfo {author} {\bibfnamefont {H.}~\bibnamefont {Ben-Hamu}}, \bibinfo
  {author} {\bibfnamefont {M.}~\bibnamefont {Nickel}}, \ and\ \bibinfo {author}
  {\bibfnamefont {M.}~\bibnamefont {Le}},\ }in\ \href@noop {} {\emph {\bibinfo
  {booktitle} {International Conference on Learning Representations (ICLR)}}}\
  (\bibinfo {year} {2023})\BibitemShut {NoStop}%
\bibitem [{\citenamefont {Klinger}\ and\ \citenamefont
  {Rotskoff}(2025)}]{Klinger2025}%
  \BibitemOpen
  \bibfield  {author} {\bibinfo {author} {\bibfnamefont {J.}~\bibnamefont
  {Klinger}}\ and\ \bibinfo {author} {\bibfnamefont {G.~M.}\ \bibnamefont
  {Rotskoff}},\ }\href@noop {} {\bibfield  {journal} {\bibinfo  {journal}
  {Phys. Rev. E}\ }\textbf {\bibinfo {volume} {111}},\ \bibinfo {pages}
  {014114} (\bibinfo {year} {2025})}\BibitemShut {NoStop}%
\bibitem [{\citenamefont {Ikeda}\ \emph {et~al.}(2025)\citenamefont {Ikeda},
  \citenamefont {Uda}, \citenamefont {Okanohara},\ and\ \citenamefont
  {Ito}}]{Ikeda2025}%
  \BibitemOpen
  \bibfield  {author} {\bibinfo {author} {\bibfnamefont {K.}~\bibnamefont
  {Ikeda}}, \bibinfo {author} {\bibfnamefont {T.}~\bibnamefont {Uda}}, \bibinfo
  {author} {\bibfnamefont {D.}~\bibnamefont {Okanohara}}, \ and\ \bibinfo
  {author} {\bibfnamefont {S.}~\bibnamefont {Ito}},\ }\href@noop {} {\bibfield
  {journal} {\bibinfo  {journal} {Phys. Rev. X}\ }\textbf {\bibinfo {volume}
  {15}},\ \bibinfo {pages} {031031} (\bibinfo {year} {2025})}\BibitemShut
  {NoStop}%
\bibitem [{Note4()}]{Note4}%
  \BibitemOpen
  \bibinfo {note} {In Eq.~\protect \eqref {eq:Wtheta}, $\nabla \cdot {\protect
  \bm {u}}_\theta =\mu T \nabla ^2 S_\theta - \mu \nabla ^2 U$.}\BibitemShut
  {Stop}%
\bibitem [{\citenamefont {Seifert}(2012)}]{seifert2012stochastic}%
  \BibitemOpen
  \bibfield  {author} {\bibinfo {author} {\bibfnamefont {U.}~\bibnamefont
  {Seifert}},\ }\href@noop {} {\bibfield  {journal} {\bibinfo  {journal} {Rep.
  Prog. Phys.}\ }\textbf {\bibinfo {volume} {75}},\ \bibinfo {pages} {126001}
  (\bibinfo {year} {2012})}\BibitemShut {NoStop}%
\bibitem [{\citenamefont {Peliti}\ and\ \citenamefont
  {Pigolotti}(2021)}]{Peliti2021}%
  \BibitemOpen
  \bibfield  {author} {\bibinfo {author} {\bibfnamefont {L.}~\bibnamefont
  {Peliti}}\ and\ \bibinfo {author} {\bibfnamefont {S.}~\bibnamefont
  {Pigolotti}},\ }\href@noop {} {\emph {\bibinfo {title} {Stochastic
  Thermodynamics: An Introduction}}}\ (\bibinfo  {publisher} {Princeton
  University Press},\ \bibinfo {year} {2021})\BibitemShut {NoStop}%
\bibitem [{\citenamefont {Limmer}(2024)}]{Limmer2024}%
  \BibitemOpen
  \bibfield  {author} {\bibinfo {author} {\bibfnamefont {D.~T.}\ \bibnamefont
  {Limmer}},\ }\href@noop {} {\emph {\bibinfo {title} {Statistical Mechanics
  and Stochastic Thermodynamics}}}\ (\bibinfo  {publisher} {Oxford University
  Press},\ \bibinfo {year} {2024})\BibitemShut {NoStop}%
\bibitem [{\citenamefont {Seifert}(2025)}]{Seifert2025}%
  \BibitemOpen
  \bibfield  {author} {\bibinfo {author} {\bibfnamefont {U.}~\bibnamefont
  {Seifert}},\ }\href@noop {} {\emph {\bibinfo {title} {Stochastic
  Thermodynamics}}}\ (\bibinfo  {publisher} {Cambridge University Press},\
  \bibinfo {year} {2025})\BibitemShut {NoStop}%
\bibitem [{\citenamefont {Anderson}(1982)}]{anderson1982reverse}%
  \BibitemOpen
  \bibfield  {author} {\bibinfo {author} {\bibfnamefont {B.~D.}\ \bibnamefont
  {Anderson}},\ }\href@noop {} {\bibfield  {journal} {\bibinfo  {journal}
  {Stoch. Process. Their Appl.}\ }\textbf {\bibinfo {volume} {12}},\ \bibinfo
  {pages} {313} (\bibinfo {year} {1982})}\BibitemShut {NoStop}%
\bibitem [{\citenamefont {Ho}\ \emph {et~al.}(2020)\citenamefont {Ho},
  \citenamefont {Jain},\ and\ \citenamefont {Abbeel}}]{ho2020denoising}%
  \BibitemOpen
  \bibfield  {author} {\bibinfo {author} {\bibfnamefont {J.}~\bibnamefont
  {Ho}}, \bibinfo {author} {\bibfnamefont {A.}~\bibnamefont {Jain}}, \ and\
  \bibinfo {author} {\bibfnamefont {P.}~\bibnamefont {Abbeel}},\ }in\
  \href@noop {} {\emph {\bibinfo {booktitle} {Advances in Neural Information
  Processing Systems (NeurIPS)}}}\ (\bibinfo {year} {2020})\BibitemShut
  {NoStop}%
\bibitem [{\citenamefont {Wang}\ \emph {et~al.}(2022)\citenamefont {Wang},
  \citenamefont {Herron},\ and\ \citenamefont {Tiwary}}]{Yihang2022Fromdata}%
  \BibitemOpen
  \bibfield  {author} {\bibinfo {author} {\bibfnamefont {Y.}~\bibnamefont
  {Wang}}, \bibinfo {author} {\bibfnamefont {L.}~\bibnamefont {Herron}}, \ and\
  \bibinfo {author} {\bibfnamefont {P.}~\bibnamefont {Tiwary}},\ }\href
  {\doibase 10.1073/pnas.2203656119} {\bibfield  {journal} {\bibinfo  {journal}
  {Proc. Natl. Acad. Sci. U.S.A.}\ }\textbf {\bibinfo {volume} {119}},\
  \bibinfo {pages} {e2203656119} (\bibinfo {year} {2022})}\BibitemShut
  {NoStop}%
\bibitem [{\citenamefont {Xu}\ \emph {et~al.}(2022)\citenamefont {Xu},
  \citenamefont {Yu}, \citenamefont {Song}, \citenamefont {Shi}, \citenamefont
  {Ermon},\ and\ \citenamefont {Tang}}]{xu2022geodiff}%
  \BibitemOpen
  \bibfield  {author} {\bibinfo {author} {\bibfnamefont {M.}~\bibnamefont
  {Xu}}, \bibinfo {author} {\bibfnamefont {L.}~\bibnamefont {Yu}}, \bibinfo
  {author} {\bibfnamefont {Y.}~\bibnamefont {Song}}, \bibinfo {author}
  {\bibfnamefont {C.}~\bibnamefont {Shi}}, \bibinfo {author} {\bibfnamefont
  {S.}~\bibnamefont {Ermon}}, \ and\ \bibinfo {author} {\bibfnamefont
  {J.}~\bibnamefont {Tang}},\ }in\ \href
  {https://openreview.net/forum?id=PzcvxEMzvQC} {\emph {\bibinfo {booktitle}
  {International Conference on Learning Representations (ICLR)}}}\ (\bibinfo
  {year} {2022})\BibitemShut {NoStop}%
\bibitem [{Note5()}]{Note5}%
  \BibitemOpen
  \bibinfo {note} {We have derived this result using standard tools of
  nonequilibrium statistical physics and stochastic thermodynamics~\cite
  {seifert2012stochastic,Peliti2021,Limmer2024,Seifert2025}. In the DM
  literature the result traces back to rigorous mathematical analysis~\cite
  {anderson1982reverse}.}\BibitemShut {Stop}%
\bibitem [{\citenamefont {Hummer}\ and\ \citenamefont
  {Szabo}(2001)}]{Gerhard2001FreeEnergy}%
  \BibitemOpen
  \bibfield  {author} {\bibinfo {author} {\bibfnamefont {G.}~\bibnamefont
  {Hummer}}\ and\ \bibinfo {author} {\bibfnamefont {A.}~\bibnamefont {Szabo}},\
  }\href {\doibase 10.1073/pnas.071034098} {\bibfield  {journal} {\bibinfo
  {journal} {Proc. Natl. Acad. Sci. U.S.A.}\ }\textbf {\bibinfo {volume}
  {98}},\ \bibinfo {pages} {3658} (\bibinfo {year} {2001})}\BibitemShut
  {NoStop}%
\bibitem [{\citenamefont {Hummer}\ and\ \citenamefont
  {Szabo}(2005)}]{HummerSzabo2005}%
  \BibitemOpen
  \bibfield  {author} {\bibinfo {author} {\bibfnamefont {G.}~\bibnamefont
  {Hummer}}\ and\ \bibinfo {author} {\bibfnamefont {A.}~\bibnamefont {Szabo}},\
  }\href@noop {} {\bibfield  {journal} {\bibinfo  {journal} {Acc. Chem. Res.}\
  }\textbf {\bibinfo {volume} {38}},\ \bibinfo {pages} {504} (\bibinfo {year}
  {2005})}\BibitemShut {NoStop}%
\bibitem [{\citenamefont {Ge}\ and\ \citenamefont {Jiang}(2008)}]{GeJiang2008}%
  \BibitemOpen
  \bibfield  {author} {\bibinfo {author} {\bibfnamefont {H.}~\bibnamefont
  {Ge}}\ and\ \bibinfo {author} {\bibfnamefont {D.-Q.}\ \bibnamefont {Jiang}},\
  }\href@noop {} {\bibfield  {journal} {\bibinfo  {journal} {J. Stat. Phys.}\
  }\textbf {\bibinfo {volume} {131}},\ \bibinfo {pages} {675} (\bibinfo {year}
  {2008})}\BibitemShut {NoStop}%
\bibitem [{Note6()}]{Note6}%
  \BibitemOpen
  \bibinfo {note} {Note that the sum ${\protect \bm {v}}_\theta + {\protect \bm
  {u}}_\theta $ describes Hamiltonian flow with friction under the Hamiltonian
  $H({\protect \bm {x}},t)$.}\BibitemShut {Stop}%
\bibitem [{\citenamefont {Glorot}\ \emph {et~al.}(2011)\citenamefont {Glorot},
  \citenamefont {Bordes},\ and\ \citenamefont {Bengio}}]{ReLU}%
  \BibitemOpen
  \bibfield  {author} {\bibinfo {author} {\bibfnamefont {X.}~\bibnamefont
  {Glorot}}, \bibinfo {author} {\bibfnamefont {A.}~\bibnamefont {Bordes}}, \
  and\ \bibinfo {author} {\bibfnamefont {Y.}~\bibnamefont {Bengio}},\ }in\
  \href {https://proceedings.mlr.press/v15/glorot11a.html} {\emph {\bibinfo
  {booktitle} {International Conference on Artificial Intelligence and
  Statistics (AISTATS)}}}\ (\bibinfo {year} {2011})\BibitemShut {NoStop}%
\end{thebibliography}%

\newpage
\appendix
\onecolumngrid

\section{Virtual escorted free energy estimation and diffusion models}
\label{append:virtual_and_diffusion}

The efficacy of virtual escorted free energy estimation (VE for short) hinges on the ability to infer, up to normalization, a time-dependent density, $\rho({\bm x},t)$, from a collection of trajectories ${\bm x}_t$.
Exactly the same task is central to {\it diffusion models}~\cite{sohl2015deep,ho2020denoising,song2021scorebased} (DM), a class of powerful generative machine learning models that have been widely used for image and audio generation, language processing, and other applications.
In the scientific context, DM have been used to infer phase transitions~\cite{Yihang2022Fromdata}, and to generate molecular conformations~\cite{xu2022geodiff}, among other applications.
The fact that inferring $\rho({\bm x},t)$ from data lies at the heart of both VE and DM is not a coincidence.
Below, we argue that both methods can be described within the same mathematical framework.
Specifically, we show that knowledge of $\nabla\ln\rho$ enables us to construct a particular decomposition, Eq.~\eqref{eq:vFdecomp}, that serves useful purposes in both contexts.

We begin, very generally, with the Langevin dynamics
\be
\label{eq:lang-app}
\dot{\bm x}_t = {\bm v}^F({\bm x}_t,t) + \sqrt{2D} \, {\bm \xi}(t) \, ,
\ee
where ${\bm v}^F({\bm x}_t,t)$ is a flow field, and ${\bm\xi}(t)$ is defined as in the main text.
At the ensemble level, these dynamics are described by the Fokker-Planck equation,
\be
\label{eq:fpe-app}
\partial_t \rho = - \nabla \cdot \left( {\bm v}^F \rho \right) + D \nabla^2 \rho 
\ee
Throughout this Appendix, we take $\rho({\bm x},t)$ to be a particular solution of this equation, for $t\in[0,\tau]$.
The initial conditions $\rho({\bm x},0)$ uniquely determine this solution.

If we know $\rho({\bm x},t)$ (perhaps only up to normalization), then we can decompose ${\bm v}^F$ as follows:
\be
\label{eq:vFdecomp}
{\bm v}^F = D \nabla\ln\rho + {\bm u} = {\bm v} + {\bm u} \, ,
\ee
where ${\bm v}({\bm x},t)$ and ${\bm u}({\bm x},t)$ are defined by Eq.~\eqref{eq:vFdecomp}.
With this decomposition, Eq.~\eqref{eq:fpe-app} becomes
\begin{equation}
\label{eq:fpe-decomp-app}
\partial_t \rho =  {\mathcal L}_t\rho - \nabla \cdot \left( {\bm u} \rho \right) \, ,
\end{equation}
where
\be
\label{eq:Ltdef}
{\mathcal L}_t (*) \equiv - \nabla\cdot ({\bm v} *) + D \nabla^2 (*) \, .
\ee
(The asterisk indicates a generic probability density, not necessarily our particular solution $\rho$.)

Translated into the notation of the main text, Eq.~\eqref{eq:vFdecomp} is the ``perfect'' decomposition of Eq.~\eqref{eq:lang} into Eq.~\eqref{eq:lang-reinterpreted},
\be
-\mu\nabla U = -\mu\nabla U_{\theta^\star} + {\bm u}_{\theta^\star} \, ,
\ee
since $-\mu\nabla U_{\theta^\star} = D\nabla\ln\rho$. 
As shown in the main text, this decomposition optimizes the convergence of Eq.~\eqref{eq:expWtheta}.
Now we show how the same decomposition arises in diffusion models.

From Eqs.~\eqref{eq:vFdecomp} and \eqref{eq:Ltdef} we have
\be
\label{eq:Ltrho=0}
{\mathcal L}_t\rho = -\nabla\cdot \left[ (D\nabla\ln\rho) \rho\right] + D \nabla^2\rho = 0\, .
\ee
Combining this result with Eq.~\eqref{eq:fpe-decomp-app}, we see that $\rho({\bm x},t)$ obeys the continuity equation,
\be
\label{eq:continuity-app}
\partial_t \rho = -\nabla\cdot ({\bm u} \rho) \, .
\ee
We introduce the notation $\tilde \rho({\bm x},t) = \rho({\bm x},\tau-t)$ to denote time-reversal.
Similar notation applies to $\tilde{\bm u}$, $\tilde{\bm v}^F$, $\tilde{\mathcal L}_t={\mathcal L}_{\tau-t}$, etc.
By Eq.~\eqref{eq:continuity-app}, $\tilde\rho({\bm x},t)$ satisfies
\begin{eqnarray}
\partial_t \tilde\rho &=& +\nabla\cdot( \tilde{\bm u} \tilde\rho) = \tilde{\mathcal L}_t \tilde\rho+\nabla\cdot( \tilde{\bm u} \tilde\rho) \nonumber \\
&=& -\nabla \cdot \left[ (\tilde{\bm v}^F-2\tilde{\bm u}) \tilde\rho \right] + D \nabla^2 \tilde\rho 
\end{eqnarray}
We used Eq.~\eqref{eq:Ltrho=0} on the first line, and Eqs.~\eqref{eq:vFdecomp} and \eqref{eq:Ltdef} to get to the second line.
Introducing
\be
\label{eq:vR}
{\bm v}^R({\bm x},t) \equiv {\bm v}^F({\bm x},\tau-t) - 2 {\bm u}({\bm x},\tau-t) \, ,
\ee
we obtain
\be
\label{eq:fpe-tilderho}
\partial_t \tilde\rho = -\nabla\cdot ({\bm v}^R\tilde\rho) + D \nabla^2 \tilde\rho \, .
\ee
At the single-trajectory level, this Fokker-Planck equation corresponds to the Langevin dynamics
\be
\label{eq:lang-vR}
\dot{\bm x}_t = {\bm v}^R({\bm x}_t,t) + \sqrt{2D} {\bm\xi}(t) \, .
\ee
Eq.~\eqref{eq:fpe-tilderho} thus implies that an ensemble of trajectories evolving under Eq.~\eqref{eq:lang-vR}, with ${\bm x}_0\sim\tilde\rho(0)$, is described by the ``time-reversed'' density $\tilde\rho({\bm x},t) = \rho({\bm x},\tau-t)$~\footnote{
We have derived this result using standard tools of nonequilibrium statistical physics and stochastic thermodynamics~\cite{seifert2012stochastic,Peliti2021,Limmer2024,Seifert2025}.
In the DM literature the result traces back to rigorous mathematical analysis~\cite{anderson1982reverse}.}.

In diffusion models, data points ${\bm x}_0^{(1)}, {\bm x}_0^{(2)}, \cdots$ are presumed to be sampled from an unknown distribution $\rho({\bm x},0)$.
Trajectories are numerically generated under Eq.~\eqref{eq:lang-app} from these initial conditions, with ${\bm v}^F$ and $D$ chosen to produce final points ${\bm x}_\tau^{(1)}, {\bm x}_\tau^{(2)}, \cdots$ described by a simple, known distribution $\rho({\bm x},\tau)$.
Typically, the Ornstein-Uhlenbeck (OU) process is used for Eq.~\eqref{eq:lang-app}, so that $\rho(\tau)$ is a Gaussian distribution and efficient learning is enabled by the closed-form OU propagator~\cite{song2021scorebased}.
If $\rho({\bm x},t)$ can accurately be learned from the trajectories, then ${\bm v}^R({\bm x},t)$ can be constructed from Eq.~\eqref{eq:vR}, via Eq.~\eqref{eq:vFdecomp}.
Then, sampling initial conditions ${\bm x}_0$ from the known distribution $\tilde\rho({\bm x},0)=\rho({\bm x},\tau)$ and evolving under Eq.~\eqref{eq:lang-vR}, one effectively arrives at a sample ${\bm x}_\tau$ from $\tilde\rho({\bm x},\tau)=\rho({\bm x},0)$.
This brief description of DM, which omits many important details, highlights the central role that the inferred density $\rho({\bm x},t)$ plays in constructing a process, Eq.~\eqref{eq:lang-vR}, that evolves a known distribution, $\rho({\bm x},\tau)$, to an unknown one, $\rho({\bm x},0)$.

\section{Virtually escorting with underdamped or Hamiltonian dynamics}
\label{append:Liouvillian}

We show how virtual escorting can be applied to underdamped Langevin dynamics and to Hamiltonian dynamics.
The state vector $\bm x$ consists of position and momentum vectors: $\bm x = (\bm q, \bm p)$.
We assume we are given a time-dependent Hamiltonian $H({\bm x},t)$.
Let $\pi_A({\bm x})$ and $\pi_B({\bm x})$ denote the equilibrium distributions associated with $H_A({\bm x})\equiv H({\bm x},0)$ and $H_B({\bm x})\equiv H({\bm x},\tau)$, at temperature $T$.
We wish to compute the corresponding free energy difference, $\Delta F = F_B-F_A$.

In the case of Hamiltonian dynamics, a trajectory ${\bm x}_t$ satisfies Hamilton's equations of motion.
In the underdamped Langevin case, there are additionally dissipative and noisy terms.
For both cases, let $\rho({\bm x},t)$ denote a probability density that describes an ensemble of trajectories, evolving under the given dynamics (Hamiltonian or underdamped Langevin), with initial conditions ${\bm x}_0$ sampled from $\pi_A$.
This density satisfies the Liouville-type equation
\begin{align}
    \partial_t \rho(\bm x,t) = \mathcal{L}(t) \rho(\bm x,t)
\label{eq:Liouville}
,\end{align}
where $\mathcal{L}(t)$ contains conservative terms derived from $H({\bm x},t)$, and -- in the Langevin case -- dissipative and diffusive terms.

Now let $\pi_\theta(\bm x,t)$ be a strictly positive probability density defined for $t\in(0,\tau)$.
As in the main text, $\theta$ represents parameters that will be optimized.
We introduce a {\it virtual} Hamiltonian,
\begin{align}
    H_\theta(\bm x, t) \equiv -T \ln {\pi_\theta (\bm x,t)} + c(t), \quad t\in (0,\tau)
\label{eq:Htheta} \, ,
\end{align}
and we impose the boundary conditions
\begin{align}
    H_\theta(\bm x,0) = H_A(\bm x)\quad, \quad H_\theta(\bm x,\tau ) =H_B(\bm x).
\end{align}

Next, suppose we construct a parameterized operator $\mathcal L_\theta(t)$ that satisfies
\begin{equation}
    \mathcal L_\theta(t) \pi_\theta({\bm x},t) = 0
    \label{eq:Ltheta-condition}
\end{equation}
at all times (an explicit construction is given in Subsection~\ref{append:decomp_momen}).
Using $\mathcal L_\theta$, we rewrite Eq.~\eqref{eq:Liouville} as follows:
\be
    \partial_t \rho  =  \mathcal{L}_\theta \rho + \left (  \mathcal{L}- \mathcal{L}_\theta \right)\rho
.\label{eq:decompose_op}\ee
If the final term in Eq.~\eqref{eq:decompose_op} can be rewritten in the form
\begin{align}
    \left(\mathcal{L}- \mathcal{L}_\theta \right)\rho = - \nabla \cdot\left ( \bm u_\theta\rho\right )
\label{eq:mid_cond_liouville} \, ,
\end{align}
where ${\bm u}_\theta({\bm x},t)$ is a vector field, and $\nabla \equiv \partial/\partial{\bm x}$ (as in the main text), then Eq.~\eqref{eq:Liouville} becomes
\begin{align}
    \partial_t \rho  =& \mathcal{L}_\theta \rho  - \nabla\cdot\left ( \bm u_\theta \rho\right)
\label{eq:virtualdecomp}\, .
\end{align} 
These dynamics can be reinterpreted as evolution under $\partial_t \rho = \mathcal L_\theta \rho $, escorted by the field~$\bm u_\theta$.
At this point the results of Ref.~\cite{escortedPRL2008} can be invoked, as in the main text.

For the decomposition given by Eq.~\eqref{eq:virtualdecomp}, we follow Ref.~\cite{escortedPRL2008} to define
\begin{eqnarray}
\label{eq:gen_w-escorted}
W_{\theta}[{\bm x}_t]
&\equiv&  \int_0^\tau dt \, \left( \frac{\partial H_\theta}{\partial t}  + {\bm u}_\theta \cdot \nabla H_\theta - T \nabla\cdot{\bm u}_\theta \right) \\
\label{eq:genescortedWork}
&=& \Delta H(\bm x_\tau, \bm x_0) + T\int_{0}^{\tau} dt \left [  \dot{\bm x}_t \circ \nabla \ln  \pi_\theta - \bm u_\theta \cdot \nabla \ln  \pi_\theta -  \nabla \cdot \bm u_\theta \right] \, ,
\end{eqnarray}
where $\Delta H \equiv H_B(\bm x_\tau) - H_A(\bm x_0)$ and $\circ$ denotes the Stratonovich product~\cite{gardiner1985handbook}.
In Eq.~\eqref{eq:genescortedWork} we used the identity
\be
\frac{d}{dt} H_\theta({\bm x}_t,t) = \partial_t H_\theta + \dot{\bm x}_t\circ \nabla H_\theta
\ee
and the relation $\nabla H_\theta = -T \nabla\ln\pi_\theta$ -- see Eq.~\eqref{eq:Htheta}.

Eq.~\eqref{eq:controlFT} then translates to the result 
\begin{align}
    \left\langle{e^{-W_\theta/T}}\right\rangle = e^{-\Delta F/T}
    \label{eq:appen_decomposEquality}
.\end{align}  
The angular brackets indicate an ensemble average over unescorted trajectories.

\subsection{Explicit Construction of ${\mathcal L}_\theta$}
\label{append:decomp_momen}
Given a choice of $\pi_\theta({\bm x},t)$, the validity of Eq.~\eqref{eq:appen_decomposEquality} for Hamiltonian or underdamped Langevin dynamics requires an operator ${\cal L}_\theta$ that satisfies Eqs.~\eqref{eq:Ltheta-condition} and \eqref{eq:mid_cond_liouville}, for some field ${\bm u}_\theta$, where ${\cal L}$ in Eq.~\eqref{eq:mid_cond_liouville} is the evolution operator governing the unescorted evolution.
We now construct ${\cal L}_\theta$ explicitly, for the underdamped Langevin case.
The Hamiltonian case will then follow immediately by setting the friction and diffusion coefficients to zero.

We assume a Hamiltonian that is the sum of kinetic and potential energy contributions:
\begin{equation}
\label{eq:kin+pot}
    H({\bm x},t) = \frac{{\bm p}^2}{2m} + V({\bm q},t) \, .
\end{equation}
An ensemble of underdamped Langevin trajectories is then described by a density $\rho({\bm x},t)$ that satisfies
\begin{align}
    \partial_t \rho &= {\mathcal L}\rho
    \equiv -\partial_{\bm q} \cdot\left(\frac{\bm p}{m} \rho \right )
    + \partial_{\bm p} \cdot \left[ (\partial_{\bm q} H) \, \rho \right] + \gamma\partial_{\bm p}\cdot\left(\frac{\bm p}{m}\rho\right) + D_p \partial_{\bm p}^2\rho
    \label{eq:underdampedF}
\end{align}
where $\gamma$ is a friction coefficient, $D_p = \gamma T$ is a momentum diffusion coefficient, and $\partial_{\bm q}$ and $\partial_{\bm p}$ denote partial derivatives with respect to position and momentum variables.
The first two terms on the right side of Eq.~\eqref{eq:underdampedF} describe Hamiltonian evolution (note that ${\bm p}/m=\partial_{\bm p}H$).
The other two describe dissipation and momentum diffusion.

For a given choice of $\pi_\theta({\bm x},t)$, we define the operator ${\mathcal L}_\theta$ in a similar manner:
\begin{equation}
    {\mathcal L}_\theta \rho \equiv -\partial_{\bm q} \cdot\left[ (\partial_{\bm p} H_\theta) \rho \right ]
    + \partial_{\bm p} \cdot \left[ (\partial_{\bm q} H_\theta) \, \rho \right] + \gamma\partial_{\bm p}\cdot\left[ (\partial_{\bm p} H_\theta)\rho \right] + D_p \partial_{\bm p}^2\rho \, ,
    \label{eq:underdampedFPo}
\end{equation}
where $(\partial_{\bm q} H_\theta,\partial_{\bm p} H_\theta) = \nabla H_\theta = -T\nabla\ln\pi_\theta$, by Eq.~\eqref{eq:Htheta}.
By inspection, we verify that Eq.~\eqref{eq:Ltheta-condition} is satisfied.

Subtracting Eq.~\eqref{eq:underdampedFPo} from Eq.~\eqref{eq:underdampedF} gives us
\begin{equation}
    \left( {\mathcal L} - {\mathcal L}_\theta \right) \rho = -\nabla\cdot\left( {\bm u}_\theta \rho \right) \, ,
\label{eq:substract}\end{equation}
where
\begin{equation}
\label{eq:utheta-underdamped}
    \bm u_\theta = \left(  \frac{\bm p}{m} -  \partial_{\bm p} H_\theta \ , \
    -\partial_{\bm q} H -\gamma \frac{\bm p}{m} + \partial_{\bm q} H_\theta + \gamma \partial_{\bm p} H_\theta  \right) \, .
\end{equation}
Thus the operator ${\cal L}_\theta$ defined by Eq.~\eqref{eq:underdampedFPo} satisfies Eqs.~\eqref{eq:Ltheta-condition} and \eqref{eq:mid_cond_liouville}, hence Eq.~\eqref{eq:appen_decomposEquality} is valid.

For the case of Hamiltonian evolution, ${\cal L}$, ${\cal L}_\theta$ and ${\bm u}_\theta$ are given exactly as in Eqs.~\eqref{eq:kin+pot}--\eqref{eq:utheta-underdamped}, except with $\gamma, D_p=0$.

\section{Alternative derivation of Eq.~\eqref{eq:expWtheta}}
\label{append:alt_der}

An ensemble of unescorted trajectories, evolving under  the Langevin equation $\dot{\bm x}_t=-\mu\nabla U + \sqrt{2D}\,{\bm\xi}$, Eq.~\eqref{eq:lang}, with ${\bm x}_0\sim\pi_A$, is described by a probability density $\rho({\bm x},t)$ that satisfies the Fokker-Planck equation
\begin{equation}
\label{eq:fpe}
    \partial_t\rho = \mu\nabla\cdot\left(\nabla U \, \rho\right) + \mu T \, \nabla^2 \rho \equiv {\mathcal L}(t) \rho
\end{equation}
(using $D=\mu T$), together with initial conditions
\begin{equation}
    \rho({\bm x},t) = \pi_A({\bm x}) \, .
\end{equation}

Formally, the density $\rho({\bm x},t)$ can be written as
\begin{equation}
    \rho({\bm x},t) = \left\langle \delta\left( {\bm x} - {\bm x}_t \right)\right\rangle \, ,
\end{equation}
where ${\bm x}_t$ is a trajectory, and $\langle\cdots\rangle$ denotes an average over an ensemble of such trajectories. 
For the same ensemble, and for a given choice of $\pi_\theta({\bm x},t)$, we define
\begin{equation}
    g({\bm x},t) \equiv \left\langle \delta\left( {\bm x} - {\bm x}_t \right) \, e^{-\beta w_{\theta,t}}\right\rangle \, .
\end{equation}
Here,
\begin{equation}
\label{eq:wtheta}
    w_{\theta,t} \equiv \int_0^t ds \,
    \left( \partial_t U_\theta + {\bm u}_\theta\cdot\nabla U_\theta - T\nabla\cdot{\bm u}_\theta \right)
\end{equation}
is the virtual work performed, {\it up to time $t$}, on the system during a realization described by the trajectory ${\bm x}_t$.
Hence, $W_\theta = w_{\theta,\tau}$ (see Eq.~\eqref{eq:Wtheta-initial}).

By the Feynman-Kac theorem~\cite{Gerhard2001FreeEnergy,HummerSzabo2005,GeJiang2008}, $g({\bm x},t)$ satisfies
\begin{equation}
\label{eq:FK}
    \partial_t g = {\mathcal L}(t) g - \beta \dot w_{\theta,t} g \, ,
\end{equation}
where $\dot w_{\theta,t}=dw_{\theta,t}/dt$.
Additionally, since $w_{\theta,0}=0$,
\begin{equation}
\label{eq:g0}
    g({\bm x},0)=\rho({\bm x},0) = \pi_A({\bm x}) \, .
\end{equation}
It is useful to think of $g({\bm x},t)$ as a {\it weighted} density, in which each trajectory carries a time-dependent statistical weight $\exp (-\beta w_{\theta, t})$.
The two terms on the right side of Eq.~\eqref{eq:FK} then describe contributions to the evolution of $g$ arising from, respectively, the evolution of the trajectories in ${\bm x}$-space, and the evolution of the statisical weights~\cite{ChrisPRE1997}.

Using Eqs.~\eqref{eq:uUUtheta}, ~\eqref{eq:fpe} and \eqref{eq:wtheta} to evaluate Eq.~\eqref{eq:FK}, we obtain
\begin{equation}
\label{eq:dgdt}
    \begin{split}
         \partial_t g &= \mu\nabla\cdot\left(\nabla U \, g\right) + \mu T \, \nabla^2 g
        -\beta \left( \partial_t U_\theta + {\bm u}_\theta\cdot\nabla U_\theta - T\nabla\cdot{\bm u}_\theta \right) g \\
        &= \left( \mu T \nabla - {\bm u}_\theta \right) \cdot \left[ e^{-\beta U_\theta} \nabla \left( e^{\beta U_\theta} g \right) \right]
        - \beta (\partial_t U_\theta) g \, .
    \end{split}
\end{equation}
By inspection, we see that the function $\exp[\beta (F_A - U_\theta)]$ is a solution of Eq.~\eqref{eq:dgdt}.
Moreover, by Eq.~\eqref{eq:bc}, this function satisfies the initial conditions Eq.~\eqref{eq:g0}.
Thus,
\begin{equation}
    \left\langle \delta\left( {\bm x} - {\bm x}_t \right) \, e^{-\beta w_{\theta,t}}\right\rangle = e^{\beta[F_A - U_\theta({\bm x},t)]} \, .
\end{equation}
Setting $t=\tau$, integrating both sides over ${\bm x}$, and again invoking Eq.~\eqref{eq:bc}, we arrive at the desired result:
\begin{equation}
    \left\langle e^{-\beta W_\theta}\right\rangle = e^{-\beta\Delta F} \, .
\end{equation}

\section{Derivation of Eq.~\eqref{eq:twoTerms} for overdamped Langevin dynamics}
\label{append:error_deriv}
We derive Eq.~\eqref{eq:twoTerms} for the overdamped Langevin dynamics given by Eqs.~\eqref{eq:lang} and \eqref{eq:lang-reinterpreted}:
\begin{align}
\label{eq:2langs}
\begin{split}
    \dot{\bm   x}_t &= -\mu \mathbf  \nabla U(\bm x_t, t) + \sqrt{2\mu T} \, \bm  \xi(t) \\
    &= -\mu \nabla U_\theta(\bm x_t,t)  + \bm  u_\theta(\bm x_t,t) +\sqrt{2\mu T}\bm  \xi(t) \, .
\end{split}
\end{align}
For the underdamped case, see Appendix~\ref{appendix:underdamped_Deriv}.

We begin by rewriting Eq.~\eqref{eq:Wtheta}, using $\nabla U_\theta=-T\nabla\ln\pi_\theta$ (see Eq.~\eqref{eq:Utheta}), as follows:
\begin{eqnarray}
W_\theta[{\bm x}_t] &=& \Delta U + \int_0^\tau dt \, \left[-(\dot{\bm x}_t-{\bm u}_\theta)\circ \nabla U_\theta  - T \nabla\cdot{\bm u}_\theta \right] \nonumber\\
&=& \Delta U + \int_0^\tau dt \, \left[ \left( \mu\nabla U_\theta - \sqrt{2\mu T} \, {\bm\xi} \right) \circ \nabla U_\theta - \mu T\, \nabla \cdot \nabla( U_\theta -U) \right] \nonumber\\
&=& \Delta U + \int_0^\tau dt \, \left[ \mu \left\vert \nabla U_\theta \right\vert^2  - \sqrt{2\mu T} \, {\bm\xi}  \cdot \nabla U_\theta - \mu T \, \nabla^2 U_\theta - \mu T \, \nabla^2( U_\theta -U) \right] \nonumber\\
&=& \Delta U + \mu T \int_0^\tau dt \, \left[ \nabla^2 U - 2\nabla^2 U_\theta + T^{-1} \left\vert \nabla U_\theta \right\vert^2 - \sqrt{2\mu T} \, {\bm\xi}  \cdot \nabla U_\theta\right] \, .
\label{eq:Wtheta-appendix}
\end{eqnarray}
We used Eqs.~\eqref{eq:uUUtheta} and \eqref{eq:2langs} to get to the second line.
To get to the third line, we used stochastic calculus to replace the Stratonovich product ($\circ$) with the It{\^ o} product ($\cdot$)~\cite{gardiner1985handbook}:
\begin{equation}
\label{eq:stochasticCalculus}
    \sqrt{2\mu T} \, {\bm\xi}\circ\nabla U_\theta =
    \sqrt{2\mu T} \, {\bm\xi}\cdot\nabla U_\theta
    + \mu T \nabla^2 U_\theta \, .
\end{equation}
Taking the ensemble average of both sides of Eq.~\eqref{eq:Wtheta-appendix}, and again using $\nabla U_\theta=-T\nabla\ln\pi_\theta$, we get
\be
\langle W_\theta\rangle = \langle\Delta U\rangle + \mu T \int_0^\tau dt \,
\left\langle \nabla^2 U + 2T \nabla^2 \ln\pi_\theta + T \left( \nabla\ln\pi_\theta \right)^2 \right\rangle
\label{eq:Wvinter}
.\ee
Integrating by parts, we obtain 
\begin{eqnarray}
\left\langle \nabla^2 \ln \pi_\theta \right\rangle &=& \int d{\bm x} \, \rho \, \nabla^2 \ln \pi_\theta
= - \int d{\bm x} \, \nabla \rho \cdot \nabla \ln \pi_\theta \nonumber\\
&=& - \int d{\bm x} \, \rho \left( \nabla \ln \rho \right) \cdot \left( \nabla \ln \pi_\theta \right) 
= - \Bigl\langle \left( \nabla \ln \rho \right) \cdot \left( \nabla \ln \pi_\theta \right) \Bigr\rangle \, ,
\end{eqnarray}
which allows us to rewrite Eq.~\eqref{eq:Wvinter} as
\be
\label{eq:Wtheta_avg}
\langle W_\theta\rangle
=  \langle\Delta U\rangle + \mu T \int_0^\tau dt \,
\left\langle \nabla^2 U + T \left( \nabla \ln \frac{\rho}{\pi_\theta} \right)^2 - T \left( \nabla\ln\rho \right)^2 \right\rangle
\ee

Next, we introduce the Shannon entropy of the ensemble of trajectories:
\be
S(t) = -\int d{\bm x} \, \rho({\bm x},t) \ln \rho({\bm x},t)
\ee
Its time derivative is
\begin{eqnarray}
\dot S &=& -\int d{\bm x} \, (\partial_t \rho) \ln \rho = -\int d{\bm x} \, \left[ \mu \nabla\cdot \left( \rho \nabla U \right) + \mu T \nabla^2 \rho \right] \ln \rho \nonumber\\
&=& \mu \int d{\bm x} \, \left[ \nabla\rho \cdot \nabla U + T \frac{ (\nabla\rho)^2 }{ \rho } \right] \nonumber\\
\label{eq:Sdot}
&=& \mu \int d{\bm x} \, \rho \left[ -\nabla^2 U + T \left( \nabla\ln\rho \right)^2 \right] \, ,
\end{eqnarray}
after integrating by parts, twice.

Eqs.~\eqref{eq:Wtheta_avg} and \eqref{eq:Sdot} give us
\be
\langle W_\theta\rangle = \left[ \langle\Delta U\rangle -T \int_0^\tau dt \, \dot S(t) \right] + \mu T^2 \int_0^\tau dt \,
\left\langle \left( \nabla \ln \frac{\pi_\theta}{\rho} \right)^2  \right\rangle
\label{eq:squared}.\ee

Recall the Kullback--Leibler and Fisher divergences between probability distributions $f$ and $g$:
\begin{equation}
{\mathcal D}_{\rm KL}[f || g] = \int f \ln \frac{f}{g}  \qquad,\qquad 
{\mathcal D}_{\rm F}[f || g] = \frac{1}{2} \int f \left( \nabla\ln\frac{f}{g} \right)^2 \, .
\end{equation}
The quantity in square brackets in Eq.~\eqref{eq:squared} can be rewritten as follows:
\begin{eqnarray}
\label{eq:noneqDeltaF}
\left[ \cdots \right] &=& \int d{\bm x} \, \rho(\tau) \left[ U_B + T \ln\rho(\tau) \right] - \int d{\bm x} \, \rho(0) \left[ U_A + T \ln\rho(0) \right] \nonumber\\
&=& \int d{\bm x} \, \rho(\tau) \left( U_B + T \ln\pi_B \right)
- \int d{\bm x} \, \pi_A \left( U_A + T \ln\pi_A \right)
+ T \, {\mathcal D}_{\rm KL} \left[ \rho(\tau) \, \vert\vert \, \pi_B \right] \nonumber\\
&=& \Delta F + T \, {\mathcal D}_{\rm KL} \left[ \rho(\tau) \, \vert\vert \, \pi_B \right] \, .
\end{eqnarray}
Additionally,
\be
\left\langle \left( \nabla \ln \frac{\pi_\theta}{\rho} \right)^2  \right\rangle = 2 {\mathcal D}_{\rm F} \left[ \rho(t) \, \vert\vert \, \pi_\theta(t) \right] \, .
\ee
Combining results, we finally arrive at Eq.~\eqref{eq:twoTerms}:
\be
W_\theta^{\rm diss} = T{\mathcal D}_{\rm KL} \left[ \rho(\tau) \, \vert\vert \, \pi_B \right] + 2DT \int_0^\tau dt \, {\mathcal D}_{\rm F} \left[ \rho(t) \, \vert\vert \, \pi_\theta(t) \right] \, .
\ee

\section{Extension of Eq.~\eqref{eq:twoTerms} to underdamped Langevin dynamics}
\label{appendix:underdamped_Deriv}

In the case of underdamped Langevin dynamics, the unescorted evolution is described, at the ensemble level, by Eq.~\eqref{eq:virtualdecomp}, with ${\bm x} = ({\bm q},{\bm p})$
At the single-trajectory level, this evolution satisfies the Langevin equation
\begin{align}
\label{eq:decomposedUnderdampedLang}
\dot{\bm  x}_t
=& \bm{v}_\theta + \bm{u}_\theta + \sqrt{2D_p}\cdot \bm{\xi}(t)
.\end{align}
Here,  $\xi_i(t) = 0$ for the position variables, $\langle \xi_i(t)\xi_j(t')\rangle = \delta_{ij}\delta(t - t')$ for the momentum variables, and
\begin{align}
\bm{v}_\theta
&= \left(
\partial_{\bm p} H_\theta \, ,
\, -\partial_{\bm q} H_\theta - \gamma \partial_{\bm p} H_\theta
\right) \\
\bm {u}_\theta
&= \left(
\partial_{\bm p} H - \partial_{\bm p} H_\theta \, , \,
 -\partial_{\bm q} H 
- \gamma \partial_{\bm p} H  +\partial_{\bm q} H_\theta
+ \gamma \partial_{\bm p} H_\theta
\right)
.\end{align}
We interpret ${\bm v}_\theta$ as Hamiltonian flow, with friction, under the virtual Hamiltonian $H_\theta({\bm x},t)$; and ${\bm u}_\theta$ as a virtual escorting field in phase space~\footnote{
Note that the sum ${\bm v}_\theta + {\bm u}_\theta$ describes Hamiltonian flow with friction under the Hamiltonian $H({\bm x},t)$.
}.

Using Eqs.~\eqref{eq:Htheta} and \eqref{eq:decomposedUnderdampedLang}, we write the expression for virtual work, Eq.~\eqref{eq:genescortedWork}, as follows:
\begin{align}
W_\theta[{\bm x}_t]
= \Delta H(\bm  x_\tau ,\bm  x_0)
+ \int_0^\tau dt \Big[
-\big(\bm {v}_\theta + \sqrt{2D_p}\,\boldsymbol{\xi}\big)\circ \nabla H_\theta 
- T \nabla\cdot \bm {u}_\theta
\Big].
\end{align}
Rewriting the Stratonovich term (see, e.g., Eq.~\eqref{eq:stochasticCalculus}), averaging over trajectories, and using $D_p=\gamma T$, we have
\begin{align}
\ave{ W_\theta} 
= \ave{\Delta H(\bm  x_\tau ,\bm  x_0) }
+ \int_0^\tau dt \ave{
- \bm {v}_\theta\cdot \nabla H_\theta
- \gamma T\partial_{\bm p}^2 H_\theta
- T\nabla\cdot \bm {u}_\theta
}
.\end{align}

From the definitions of $\bm v_\theta$ and $\bm u_\theta$ we obtain, after cancellations:
\begin{align}
    -\bm {v}_\theta\cdot \nabla H_\theta
    &= (-\partial_{\bm p} H_\theta \, , \, 
     \partial_{\bm q} H_\theta
    + \gamma \partial_{\bm p} H_\theta)\cdot (\partial_{\bm q} H_\theta , \partial_{\bm p} H_\theta)  
    = \gamma \, \vert\partial_{\bm p} H_\theta\vert^2 \\
    \nabla\cdot{\bm u}_\theta
    &= 
    (\partial_{\bm q} \, ,\,\partial_{\bm p}) \, \cdot (\partial_{\bm p}H
    - \partial_{\bm p} H_\theta \, , \, 
    -\partial_{\bm q} H
    - {\gamma}\partial_{\bm p} H
    + \partial_{\bm q} H_\theta
    + \gamma \partial_{\bm p} H_\theta)
    = \gamma \partial_{\bm p}^2(H_\theta-H) \, .
\end{align}
Substituting these results into our expression for $\ave{W_\theta}$, we get
\begin{eqnarray}
\ave{ W_\theta} 
&=& \left \langle \Delta H  \right\rangle
+ \gamma T\int^\tau_0 dt \, \left\langle{
T^{-1} |\partial_{\bm p} H_\theta|^2
-2 \partial_{\bm p}^2 H_\theta
+\partial_{\bm p}^2 H
}\right\rangle \\
&=& \ave{\Delta H}
+ \gamma T^2\int_0^\tau dt \,\left\langle
 |\partial_{\bm p} \ln{ \pi_\theta}|^2
+2 \partial_{\bm p}^2 \ln{ \pi_\theta}
+T^{-1} \partial_{\bm p}^2 H
\right\rangle  \\
&=& \left\langle \Delta H \right\rangle
+ \gamma T^2\int_0^\tau dt \, \left\langle
 |\partial_{\bm p} \ln{ \pi_\theta}|^2
-2 \partial_{\bm p}\ln{\rho} \cdot \partial_{\bm p} \ln{ \pi_\theta}
+T^{-1} \partial_{\bm p}^2 H \right\rangle \\
&=& 
\label{eq:Wvint}
\left\langle \Delta H \right\rangle
+ \gamma T\int_0^\tau dt \, \left\langle
 T|\partial_{\bm p} \ln{( \pi_\theta/\rho )}|^2 - T|\partial_{\bm p} \ln \rho|^2
+ \partial_{\bm p}^2 H
\right\rangle
\end{eqnarray}
using Eq.~\eqref{eq:Htheta} to get to the second line, and integration by parts to get to the third line.

Next, we take the time derivative of the system entropy, $S=-\int d{\bm x}\rho\ln\rho$, as follows:
\begin{align}
\dot S =& -\int d\bm x\, \dot\rho\ln\rho\nonumber\\
=& - \int d\bm x \, \ln \rho
\left[
- \partial_{\bm q} \left[\left(\partial_{\bm p} H\right) \, \rho \right]
+ \partial_{\bm p} \left[\left(\gamma \partial_{\bm p} H + \partial_q H\right)\rho \right]
+ D_p \partial_{\bm p}^2 \rho
\right] \, ,
\end{align}
using Eq.~\eqref{eq:underdampedF}.
Integrating by parts, we rewrite the three terms on the right as follows (here, $\int = \int d{\bm x}$):
\begin{eqnarray}
-\int  \ln \rho \, \partial_{\bm q} [(\partial_{\bm p} H) \rho]
&=& \int  \partial_{\bm q} (\ln \rho)\, (\partial_{\bm p} H ) \rho 
=- \int  \rho \, \partial_{\bm q}\partial_{\bm p} H  \\
\int \ln \rho \, \partial_{\bm p} \bigl[(\gamma \partial_{\bm p} H + \partial_{\bm q} H)\rho\bigr]
&=& - \int  \, (\partial_{\bm p}\rho)\, [(\gamma \partial_{\bm p} H + \partial_{\bm q} H)] 
= \int   \rho\, [(\gamma \partial_{\bm p}^2 H + \partial_{\bm q}\partial_{\bm p} H)] \\
\int  \ln \rho \, \partial_{\bm p} D_p \partial_{\bm p} \rho
&=& - \int  \, \partial_{\bm p} (\ln \rho)\, D_p \partial_{\bm p} \rho
= - \int \rho \, D_p \bigl|\partial_{\bm p} \ln \rho\bigr|^2
.\end{eqnarray}
Combining these results, and using $D_p=\gamma T$, we obtain
\begin{align}
\label{eq:Sdotfinal}
\dot S = - \gamma \int d\bm x \, \rho \, [ \partial_{\bm p}^2 H - T |\partial_{\bm p} \ln \rho|^2 ] \, .
\end{align}

The two terms appearing in the integrand in Eq.~\eqref{eq:Sdotfinal} also appear in Eq.~\eqref{eq:Wvint}, allowing us to write
\begin{align}
\langle W_\theta \rangle
=& \langle \Delta H \rangle
- T \int_0^\tau dt \, \dot S(t) + \gamma T^2
\int_0^\tau dt \left\langle \bigl|\partial_{\bm p} \ln{(\pi_\theta/\rho)} \bigr|^2
\right\rangle \, .
\end{align}
Taking the same steps as in Eq.~\eqref{eq:noneqDeltaF}, the first two terms on the right in the above equation can rewritten as
\begin{align}
    \Delta F + T \, {\mathcal D}_{\rm KL} \left[ \rho(\tau) \, \vert\vert \, \pi_B \right] \, .
\end{align}
We finally arrive at the following expression for dissipated work: 
\begin{align}
\label{eq:Wdiss-underdamped}
    W_{\rm diss} =  T\, {\mathcal D}_{\rm KL} \left[ \rho(\tau) \, \vert\vert \, \pi_B \right]
+ \gamma T^2
\int_0^\tau dt \int d{\bm x} \, \rho \, \left|\partial_{\bm p} \ln \left( \frac{\pi_\theta}{\rho} \right) \right|^2
\ .\end{align}

Eq.~\eqref{eq:Wdiss-underdamped} is nearly identical to Eq.~\eqref{eq:twoTerms} of the main text.
Notice, however, that integrand appearing in Eq.~\eqref{eq:Wdiss-underdamped} involves derivatives with respect to momentum, but not position.
This difference prevents us from writing the last term as the time integral of a Fisher divergence.
Additionally, it implies that the optimality condition in the underdamped case is looser than Eq.~\eqref{eq:optimizationCondition}.
Specifically, if we set
\begin{equation}
\label{eq:factorize}
    \pi_\theta({\bm q},{\bm p},t) \propto f({\bm q},t) \, \rho({\bm q},{\bm p},t) \, ,
\end{equation}
for any $f({\bm q},t)>0$, then the final term in Eq.~\eqref{eq:Wdiss-underdamped} vanishes, and $W_{\rm diss}$ is minimized.
I.e.\ to optimize the convergence of Eq.~\eqref{eq:appen_decomposEquality}, we need to determine $\rho({\bm x},t)$ only up to a multiplicative function of ${\bm q}$ and $t$.
[In the overdamped case, we need to determine $\rho({\bm x},t)$ up to a multiplicative function of $t$.]

If Eq.~\eqref{eq:factorize} is satisfied and if, additionally, the system ends in equilibrium, i.e.\ $\rho(\tau)=\pi_B$, then $W_{\rm diss}=0$, hence $W_\theta[{\bm x}_t]=\Delta F$ for every trajectory.

\section{Cost function $L(\theta)$ for minimizing the Fisher term}
\label{append:L_Fisher}

We show that minimizing $L(\theta)$, given by Eq.~\eqref{eq:costFunction}, is equivalent to minimizing the Fisher term in Eq.~\eqref{eq:twoTerms}.

We rewrite the cost function as
\begin{equation}
\label{eq:cost-SI}
    L(\theta) = \int_0^\tau dt \, \ell[\pi_\theta(t),t] \, ,
\end{equation}
where
\begin{equation}
\begin{split}
    \label{eq:ell}
    \ell[\pi_\theta(t),t] &= \int d{\bm x} \, \rho({\bm x},t) \left [|\nabla S_\theta (\bm x,t) |^2 -2\nabla^2 S_\theta (\bm x,t)  \right] \\
    &= \int d{\bm x} \, \rho \left[ \vert\nabla\ln\pi_\theta \vert^2 + 2 \nabla^2 \ln\pi_\theta \right] \\
    &= \int d{\bm x} \, \left[ \rho \vert\nabla\ln\pi_\theta \vert^2 - 2 \nabla\rho \cdot \nabla\ln\pi_\theta \right] \\
    &= \int d{\bm x} \, \rho \left[ \vert\nabla\ln\pi_\theta \vert^2 - 2 \nabla\ln\rho \cdot \nabla\ln\pi_\theta  \right] \\
    &= 2{\cal D}_F[ \rho(t) \vert\vert \pi_\theta(t) ] - \int d{\bm x} \, \rho\, |\nabla\ln\rho|^2 \, .
\end{split}
\end{equation}
We have used $\pi_\theta\propto e^{-S_\theta}$ (see text after Eq.~\eqref{eq:costFunction}) to get from the first to the second line; integration by parts to get to the third line; and
\begin{equation}
    {\cal D}_F[f||g] \equiv \frac{1}{2}\int f \,  \left( \nabla\ln\frac{f}{g}\right)^2
\end{equation}
to arrive at the last line.
The last term on the last line, $-\int \rho\, |\nabla\ln\rho|^2$, does not depend on $\theta$.
As a result, by minimizing $\ell[\pi_\theta(t),t]$ with respect to the parameters $\theta$, we minimize ${\cal D}_F[ \rho(t) \vert\vert \pi_\theta(t) ]$, at any time $t$.
Since $L(\theta)$ is the time integral of $\ell[\pi_\theta(t),t]$, we conclude that by minimizing $L(\theta)$ we minimize the Fisher term,
\begin{equation}
    2DT \int_0^\tau dt \, {\mathcal D}_{\rm F} \left[ \rho(t) \, \vert\vert \, \pi_\theta(t) \right] \, ,
\end{equation}
that appears in Eq.~\eqref{eq:twoTerms}.

Recall that our original goal was to minimize $W_\theta^{\rm diss}$, in order to optimize the convergence of Eq.~\eqref{eq:expWtheta}.
From Eqs.~\eqref{eq:twoTerms}, \eqref{eq:cost-SI} and \eqref{eq:ell} we have
\begin{equation}
\label{eq:Wtheta_Ltheta}
    W_\theta^{\rm diss} = DT L(\theta) + [\textrm{terms that do not depend on } \theta] \, ,
\end{equation}
which shows that we minimize $W_\theta^{\rm diss}$ by minimizing $L(\theta)$.
In fact, Eq.~\eqref{eq:Wtheta_Ltheta} can straightforwardly be derived directly from the definition of $W_\theta^{\rm diss}$ (details not shown), without deriving Eq.~\eqref{eq:twoTerms} along the way.
Nevertheless, Eq.~\eqref{eq:twoTerms} is useful as it identifies the dissipated work as a sum of contributions associated with: (1) the final state $\rho({\bm x},\tau)$ (the KL term), and (2) the choice of $\pi_\theta({\bm x},t)$ (the Fisher term).
Contribution (1) can be addressed by including a relaxation stage in our protocol, as discussed in the main text; and (2) is addressed by minimizing $L(\theta)$.

\section{$W_\theta$ for perfect virtual escorting}
\label{append:perfect}

Assume perfect virtual escorting:
\begin{align}
    \pi_{\theta^\star} (\bm x,t) = \rho(\bm x,t) \quad \forall \quad t\in (0,\tau)
    \label{eq:optimal_cond} \, .
\end{align} 
We show that, in this case, the escorted work depends only on the final microscopic state, $\bm x_\tau$, along the trajectory.

We first use Eq.~\eqref{eq:optimal_cond} to rewrite Eq.~\eqref{eq:Wtheta} (or Eq.~\eqref{eq:genescortedWork}) as
\begin{align}
    W_{\theta^\star}[{\bm x}_t] = \Delta U + T\int_{0}^{\tau} dt \,  (\dot{\bm x}_t \circ \nabla \ln\rho - {\bm u}_{\theta^\star} \cdot \nabla \ln  \rho -  \nabla \cdot \bm u_{\theta^\star} ) 
\label{eq:Wstarappend}\, .
\end{align}
Using $\partial_t \ln \rho  =  -\rho^{-1} \nabla( \bm u_{\theta^\star} \rho)$, which follows from Eq.~\eqref{eq:optimal_cond} and Eq.~\eqref{eq:mid_cond_liouville}, we rewrite Eq.~\eqref{eq:Wstarappend} as 
\begin{eqnarray}
    W_{\theta^\star}[{\bm x}_t] &=& \Delta U + T\int_{0}^{\tau} dt \, (  \dot{\bm x}_t \circ \nabla \ln  \rho +\partial_t \ln \rho ) \nonumber\\
    &=& U_B({\bm x}_\tau) - U_A({\bm x}_0) + T\int_{0}^{\tau} dt \, \frac{d}{dt} \ln\rho({\bm x}_t,t) \nonumber\\
    &=& U_B({\bm x}_\tau) + T\ln\rho({\bm x}_\tau,\tau) - F_A \, ,
    \label{eq:Wstarappend_fin}
\end{eqnarray}
using $\rho({\bm x}_0,0) = \pi_A({\bm x}_0)$.
Equivalently, we can write Eq.~\eqref{eq:Wstarappend_fin} as $ W_{\theta^\star} =T\ln{ [\rho(\bm x_\tau, \tau)/ \pi_B (\bm x_\tau) ]} +\Delta F$.

Eq.~\eqref{eq:Wstarappend_fin} holds for overdamped, underdamped, and deterministic dynamics, but with $U_B$ replaced by $H_B$ in the latter two cases.

\section{Probability density for the harmonic oscillator model}
\label{append:prop_harmonic}

An ensemble of trajectories evolving under Eq.~\eqref{eq:solvableLE} is described by a probability density $\rho(x,t)$ that obeys
\begin{equation}
\label{eq:fpe-ho}
    \partial_t \rho = \mu k(t) \partial_x( x\rho ) + \mu T \partial_x^2\rho \, .
\end{equation}
If the initial state is Gaussian with zero mean,
it follows from Eq.~\eqref{eq:fpe-ho} that $\rho(x,t)$ remains Gaussian at all times, with zero mean and with a variance that satisfies
\begin{equation}
    \frac{d}{dt} \sigma^2(t) = 2\mu T - 2\mu k(t) \sigma^2(t) \, .
\end{equation}
The general solution of this equation is
\begin{equation}
\label{eq:variance-general}
    \sigma^2(t) = \sigma^2(0) \exp \left[-\int_0^t ds\, 2\mu k(s) \right]
    + 2\mu T \int_0^t ds \exp\left[-\int_s^t ds^\prime 2\mu k(s^\prime)\right] \, .
\end{equation}

For the protocol specified in the main text, see Eq.~\eqref{eq:springprotocol}, and for a variance $\sigma^2(0)=T/k_i$ corresponding an initial equilibrium state, we can evaluate the integrals appearing in Eq.~\eqref{eq:variance-general}.
Introducing the dimensionless parameter $\alpha \equiv 2 \mu k_i k_f \tau_s/(k_f - k_i)$, and assuming $\alpha\ne 1$, we obtain the following result for the switching stage:
\begin{equation}
\label{eq:exactVariance-switching}
    \sigma^2(t) = \frac{T}{k(t)} \frac{\alpha - [k_i/k(t)]^{\alpha-1}}{\alpha-1} \quad , \quad t\in[0,\tau_s] \, .
\end{equation}
The limit $\alpha\rightarrow 0$ describes a sudden change of the spring constant, in which case the variance remains constant during the (infinitely brief) switching stage.
In the opposite limit, $\alpha\rightarrow {\rm sign}(k_f-k_i)\cdot\infty$, Eq.~\eqref{eq:exactVariance-switching} gives $\sigma^2(t)=T/k(t)$, which describes a system in quasistatic equilibrium.

For the relaxation stage, we have
\begin{align}
\label{eq:exactVariance-relaxation}
    \nonumber
    \sigma^2(t) &= \frac{T}{k_f} + \left[ \sigma^2(\tau_s) - \frac{T}{k_f} \right] e^{-2\mu k_f(t-\tau_s)} \\
    &= \frac{T}{k_f} \left[ 1 + \frac{1 - (k_i/k_f)^{\alpha-1}}{\alpha-1} e^{-2\mu k_f(t-\tau_s)} \right]
    \quad,\quad t \in (\tau_s,\tau] \, .
\end{align}

When $\alpha=1$, Eqs.~\eqref{eq:exactVariance-switching} and \eqref{eq:exactVariance-relaxation} simplify to the expression
\begin{equation}
    \sigma^2(t) = \frac{T}{k(t)} \left[ 1 - g(t)\ln \frac{k_i}{k(t)} \right] \, ,
\end{equation}
where $g(t) = 1$ during the switching stage, and $g(t) = e^{-2\mu k_f(t-\tau_s)}$ during the relaxation stage.
This result follows from Eq.~\eqref{eq:variance-general}, or from the evaluation of Eqs.~\eqref{eq:exactVariance-switching} and \eqref{eq:exactVariance-relaxation} using l'H\^{o}pital's Rule.

\section{Numerical implementation with a neural network}
\label{append:num_imp}

We describe how we constructed the estimates $\Delta F^{\rm est}$ for the quartic potential model, $U(x,\lambda)=x^4-16(1-\lambda)x^2$, when using a neural network to represent $S_\theta(x,t)$.

To optimize the parameter set $\theta$, we let $\{ t_k \}_{k=0}^K$ denote a set of equally spaced times, $t_k =k\tau/K$, from $0$ to $\tau$.
We divided the Langevin trajectories $\{ { x}_t^n \}_{n=1}^N$ generated using Eq.~\eqref{eq:lang} into training and validation sets, containing $N_{\rm tr}$ and $N_{\rm val}$ trajectories, respectively. 
We defined a cost function
\be
C(\theta) = \frac{1}{N^\prime} \frac{\tau}{K} \sum_{n=1}^{N^\prime} \sum_{k=0}^{K-1} \left[ \left\vert \nabla S_\theta({ x}_{t_k}^{n},t_k) \right\vert^2 + 2 \nabla^2 S_\theta({ x}_{t_k}^{n},t_k) \right] \, ,
\ee
which can be evaluated using a subset (batch) of $N^\prime$ trajectories, taken from either the training set or the validation set.
For large $N^\prime$ and $K$, $C(\theta)$ approximates the true cost function, $L(\theta)$ (Eq.~\eqref{eq:costFunction}).

Next, we represented the function $S_\theta({\bm x},t)$ as a neural network consisting of four layers, with $128$ units per layer.
We used the smooth GELU~\cite{hendrycks2016gaussian} activation function, $A_{\rm G}(x)  = 0.5 x[ 1 + {\rm erf}( {x}/{\sqrt{2}} ) ]$, to ensure the correct evaluation of our cost function $C$.
By contrast, the widely used ReLU~\cite{ReLU} activation function, $ A_{\rm R}(x) = 0.5 (x +|x|)$, is piecewise linear, and therefore cannot represent a non-vanishing $\nabla^2 S_\theta$.

We then trained the neural network by minimizing $C(\theta)$.
90~$\%$ of the generated trajectories were used for training; the remaining 10~$\%$ were reserved for validation. The input data for $S_\theta$ were prepared as pairs  $(x_t,t)$, which were divided into batches of size 4096.
In the optimization procedure, each batch was used to generate one update of the parameters $\theta$, and the optimization was performed for 250 epochs, with each epoch corresponding to a single pass through all of the batches.
We adopted the ADAM optimizer~\cite{kingma2015adam}. We set the learning rate, which controls the size of parameter updates, to $5\times10^{-5}$, and the weight decay, which prevents overfitting, to $10^{-4}$. 
The parameters~$\theta$ were updated by gradient-based optimization, using the cost function $C(\theta)$ evaluated on the training set.
The validation loss -- that is, the cost function evaluated on the validation set -- was recorded after each epoch. After completing the training procedure, we set $\theta^\star$ to the parameter values that gave the lowest recorded validation loss.

Having optimized the parameters as just described, we used all $N$ trajectories -- belonging to both the training and validation sets -- to construct the estimate,
\be
\label{eq:DeltaFeststar}
\Delta F_{\theta^\star}^{\rm est} = -T \ln \left( \frac{1}{N} \sum_n e^{-W_{\theta^\star}[{ x}_t^n]} \right) \, ,
\ee
with
\begin{align}
W_{\theta^\star}[{ x}_t] &= \Delta U({ x}_0,{ x}_\tau) + T \int_0^\tau dt \, \left( \dot{ x}_t \circ \nabla S_{\theta^\star} - { u}_{\theta^\star} \cdot \nabla S_{\theta^\star} + \nabla\cdot{ u}_{\theta^\star} \right) \\
{ u}_{\theta^\star}({ x},t) &= \mu \left( -T \nabla S_{\theta^\star} - \nabla U \right) \, .
\end{align}
To quantify the uncertainty in $\Delta F_{\theta^\star}^{\rm est}$, using the bootstrap method~\cite{Efron1982}, we evaluated the standard deviation of $\Delta F_{\theta^\star}^{\rm est}$ over 10000 sets of trajectories. Each set was generated by sampling $N$ trajectories, with replacement, from the entire trajectory set.

The optimization condition, Eq.~\eqref{eq:optimizationCondition}, is equivalent to the condition
\begin{equation}
\label{eq:newOC}
    -\nabla S_{\theta^\star} = \nabla\ln\rho \, .
\end{equation}
Our procedure for obtaining $\theta^\star$ does not guarantee that $S_{\theta^\star}$ perfectly satisfies Eq.~\eqref{eq:newOC}, for three reasons.
First, for a finite parameter set, the space of functions represented by $S_\theta$ typically does not contain one that satisfies Eq.~\eqref{eq:newOC} exactly.
Next, for finite $N^\prime$ and $K$, $C(\theta)$ is only approximately equal to $L(\theta)$.
Finally, the minimization routine typically does not converge to the exact minimum of $C(\theta)$ in a finite number of computational steps.
However, even for an ``imperfect'' parameter set $\theta^\star$, Eq.~\eqref{eq:expWtheta} guarantees that $e^{-W_{\theta^\star}/T}$ is an unbiased estimator of $e^{-\Delta F/T}$, hence the estimate in Eq.~\eqref{eq:DeltaFeststar} converges  to $\Delta F$ as $N\rightarrow\infty$.

\section{Fitting $\rho(x,t)$ for the quartic model to a sum of two Gaussians}
\label{append:two_Gauss}

In the main text, we considered overdamped Langevin trajectories, at temperature $T=1$, evolving in the potential
\begin{equation}
\label{eq:quarticUdef-SI}
    U(x,\lambda) = x^4 - 16(1-\lambda)x^2 \, ,
\end{equation}
as $\lambda$ is varied according to the protocol
\begin{equation}
    \lambda_t =
    \begin{cases}
    t/\tau_s  & ,\quad 0\le t\le\tau_s \\
    1  & ,\quad \tau_s\le t\le\tau
    \end{cases}
    \qquad ,
\end{equation}
with $\tau = \tau_s+0.8$.

Let the distribution $\rho(x,t)$ describe an ensemble of these trajectories, with $\rho(x,0)=\pi_A(x)\propto e^{-U(x,0)/T}$.
From the reflection symmetry of $U(x,\lambda)$ and $\rho(x,0)$, it follows that $\rho(x,t)=\rho(-x,t)$ during the entire process.
Since $U(x,0)$ forms a double-well potential with minima at $x=\pm 8$, separated by a barrier of height $64$ at the origin, $\rho(x,0)$ is a sum of two well-separated, nearly Gaussian distributions.
As $\lambda$ varies in time from 0 to 1, the two modes of the distribution approach one another, and ultimately relax toward the unimodal distribution $\pi_B \propto e^{-x^4/T}$.
With this picture in mind, we model $\rho(x,t)$ as a sum of two Gaussians, throughout the process:
\begin{subequations}
\label{eq:sum2gaussians}
\begin{eqnarray}
    \rho(x,t) &\approx& \pi_\theta(x,t) = g(x\,;\,\bar x_t,\sigma_t) \\
    g(x\,;\,\bar x,\sigma) &\equiv& \left[ e^{-(x-\bar x)^2/2\sigma^2} + e^{-(x+\bar x)^2/2\sigma^2} \right] / \sqrt{8\pi\sigma^2} \, .
\end{eqnarray}
\end{subequations}
The parameters $\theta$ specify the time-dependence of $\bar x_t$ and $\sigma_t$.
Eq.~\eqref{eq:sum2gaussians} is an excellent approximation at $t=0$, but worsens as the barrier between the two wells decreases.

By direct evaluation, we obtain the following expressions for the second and fourth moments of $g(x\,;\,\bar x,\sigma)$:
\begin{align}
    \langle x^2\rangle &= \int dx \, g \, x^2 = \bar x^2 + \sigma^2 \\
    \langle x^4\rangle &= \int dx \, g \, x^4 = \bar x^4 + 6\bar x^2\sigma^2 + 3\sigma^4 \, .
\end{align}
Inverting these equations gives
\begin{align}
\label{eq:mu_sigma}
    \bar x^4 &= \frac{3}{2} \langle x^2\rangle^2 - \frac{1}{2} \langle x^4\rangle \\
    \sigma^2 &= \langle x^2\rangle - \sqrt{\frac{3}{2} \langle x^2\rangle^2 - \frac{1}{2} \langle x^4\rangle} \, .
\end{align}

We chose thirteen values of $\tau_s$, ranging from $0.001$ to $1.0$.
For each $\tau_s$, we carried out the following steps.

First, we used Eq.~\eqref{eq:lang} to generate 1000 unescorted trajectories, %
saving data at every time step $\delta t = 0.001$, with $\mu, D=1$.
These simulations produced the data set $\{x_{t_k}^n\}$, where $n= 1, \cdots 1000$; $k=0, \cdots K=\tau/\delta t$; and $t_k=k\tau/K$.
At every time step, we used our trajectories to evaluate $\langle x^2\rangle$ and $\langle x^4\rangle$, then we used Eq.~\eqref{eq:mu_sigma} to assign values of $\bar x$ and $\sigma$ for that time step.
In this manner, we generated parameter values $\theta = \{ \bar x_t, \sigma_t \}$ that are consistent (under the approximation $\rho\approx\pi_\theta$) with the time-dependent second and fourth moments observed in the data.

Next, for each of the 1000 trajectories, we evaluated
\begin{equation}
\label{eq:Wtheta-SI}
    W_\theta[x_t] = \Delta U + T \int_0^\tau dt \left[ \left( \dot x_t-u_\theta\right) \circ \nabla\ln\pi_\theta - \nabla\cdot u_\theta \right]
\end{equation}
(see Eq.~\eqref{eq:Wtheta}), with
\begin{eqnarray}
    \nabla\ln\pi_\theta(x,t) &=& -\frac{1}{\sigma_t^2}
    \left[ x - \bar x_t \tanh\left(\frac{\bar x_t x}{\sigma_t^2}\right) \right] \\
    u_\theta(x,t) &=& -\mu \left[ 4x^3 - 32(1-\lambda_t) x \right] + \frac{D}{\sigma_t^2} \left[ x - \bar x_t \tanh\left(\frac{\bar x_t x}{\sigma_t^2}\right) \right] \\
    \nabla\cdot u_\theta(x,t) &=& -\mu \left[ 12x^2 - 32(1-\lambda_t) \right] + \frac{D}{\sigma_t^2} - \frac{\bar x_t^2 D}{\sigma_t^4} \sech^2 \left( \frac{\bar x_t x}{\sigma_t^2} \right) \, .
\end{eqnarray}
These expressions follow from Eq.~\eqref{eq:sum2gaussians}, along with Eqs.~\eqref{eq:uthetaUthetadefs} and \eqref{eq:quarticUdef-SI}.
In evaluating Eq.~\eqref{eq:Wtheta-SI}, we used mid-point discretization to implement the Stratonovich integral:
\begin{equation}
    \int_0^\tau dt \, \dot x_t \circ \nabla\ln\pi_\theta \rightarrow \sum_{k=0}^{K-1} \left( \frac{ x_{{k+1}}-x_{k} }{2} \right)
    \left[\frac{\nabla\ln\pi_\theta(x_{{k}},t_k) + \nabla\ln\pi_\theta(x_{{k+1}},t_{k+1})}{2}\right] \, .
\end{equation}
From the resulting work values $\{ W_\theta^n \}$, we computed $\Delta F_\theta^{\rm est}$ using Eq.~\eqref{eq:estimator}.

\section{Derivation of Eq.~\eqref{eq:continuity}}
\label{sec:optim_utheta}
We derive the optimality condition for the flow field ${\bm u}_\theta({\bm x},t)$, from the optimality condition for the density $\pi_\theta({\bm x},t)$.

The optimal density $\pi_{\theta^\star}$ satisfies Eq.~\eqref{eq:optimizationCondition}:
\begin{equation}
    \pi_{\theta^\star}({\bm x},t) = \rho({\bm x},t) \, .
\end{equation}
This condition combines with Eq.~\eqref{eq:uthetaUthetadefs} to give
\begin{equation}
    \mu\nabla U = -\mu T\nabla\ln\rho - {\bm u}_{\theta^\star} \, .
\end{equation}
Using this result, we obtain Eq.~\eqref{eq:continuity} from Eq.~\eqref{eq:fokker-planck}:
\begin{align}
    \partial_t\rho &= \nabla\cdot \left[ \left( -\mu T \nabla\ln\rho - {\bm u}_{\theta^\star} \right) \rho \right] + D \nabla^2 \rho \\
    &= -\mu T \nabla\cdot(\nabla\rho) - \nabla\cdot({\bm u}_{\theta^\star}\rho) + D \nabla^2 \rho \\
    &= - \nabla\cdot({\bm u}_{\theta^\star}\rho) \, ,
\end{align}
since $D=\mu T$.

\section{Avoiding evaluation of the Laplacian $\nabla^2 S_\theta$ in Eqs.~\eqref{eq:Wtheta} and \eqref{eq:costFunction}}
\label{append:avoid}

We show how $W_\theta[{\bm x}_t]$ and $L(\theta)$ can be computed without explicitly evaluating $\nabla^2 S_\theta$.

From Eqs.~\eqref{eq:lang} and \eqref{eq:stochasticCalculus} (and using $S_\theta=U_\theta/T$), we have~\cite{gardiner1985handbook}
\begin{equation}
\label{eq:stochasticCalculus-S}
    \dot{\bm x}_t \circ\nabla S_\theta =
    \dot{\bm x}_t \cdot\nabla S_\theta
    + \mu T \nabla^2 S_\theta \, .
\end{equation}
We now rewrite the integrand in Eq.~\eqref{eq:Wtheta} as follows:
\begin{eqnarray}
    (\dot{\bm x}_t-{\bm u}_\theta) \circ \nabla\ln\pi_\theta - \nabla\cdot{\bm u}_\theta
    &=& -\dot{\bm x}_t \circ \nabla S_\theta - \mu\nabla (U-TS_\theta)\cdot\nabla S_\theta + \mu \nabla^2(U-TS_\theta) \nonumber\\
    &=& -2\dot{\bm x}_t \circ \nabla S_\theta + \dot{\bm x}_t \cdot \nabla S_\theta -\mu\nabla U\cdot\nabla S_\theta + \mu T(\nabla S_\theta)^2
    + \mu\nabla^2 U \, ,
\end{eqnarray}
using Eq.~\eqref{eq:uthetaUthetadefs} on the first line; then subtracting the left side of Eq.~\eqref{eq:stochasticCalculus-S}, and adding the right side, to get to the second line.
We therefore have
\begin{equation}
    W[{\bm x}_t] = \Delta U + T \int_0^\tau dt \, \left[
    -2\dot{\bm x}_t \circ \nabla S_\theta + \dot{\bm x}_t \cdot \nabla S_\theta -\mu\nabla U\cdot\nabla S_\theta + \mu T(\nabla S_\theta)^2
    + \mu\nabla^2 U \, ,
    \right]
\end{equation}
which no longer involves the Laplacian $\nabla^2 S_\theta$.
For a time-discretized trajectory, with $K$ equal time steps from $t=0$ to $\tau$, the two terms involving $\dot{\bm x}_t$ are evaluated as follows:
\begin{align}
    dt \, \dot{\bm x}_t \circ \nabla S_\theta &\rightarrow \delta{\bm x}_k \cdot \frac{\left( \nabla S_{\theta,k} + \nabla S_{\theta,k+1} \right)}{2} \\
    dt \, \dot{\bm x}_t \cdot \nabla S_\theta &\rightarrow \delta{\bm x}_k \cdot \nabla S_{\theta,k} \quad ,
\end{align}
where, for $0\le k<K$,
\begin{align}
    \label{eq:deltax_kdef}
    \delta{\bm x}_k &= {\bm x}(t_{k+1}) - {\bm x}(t_k) \\
    \label{eq:nablaS_kdef}
    \nabla S_{\theta,k} &= \nabla S_\theta({\bm x}(t_k),t_k) \\
    t_k &= k\tau/K \, .
\end{align}

In Eq.~\eqref{eq:costFunction}, we rewrite the quantity $\nabla^2 S_\theta$ appearing in the integrand, as follows:
\begin{equation}
    \nabla^2 S_\theta = \frac{1}{\mu T} \left( \dot{\bm x}_t \circ \nabla S_\theta - \dot{\bm x}_t \cdot \nabla S_\theta \right) \, .
\end{equation}
Replacing the integral over $\rho({\bm x},t)$ in Eq.~\eqref{eq:costFunction} by a sum over $N$ time-discretized trajectories, we obtain
\begin{equation}
    \label{eq:costFunction-noLaplacian}
    L(\theta) \rightarrow \frac{1}{N} \sum_{n=1}^N \sum_{k=0}^{K-1}
    \left[
    \frac{\tau}{K} \left(\nabla S_{\theta,k}^n\right)^2 - \frac{1}{\mu T} \delta {\bm x}_k^n \cdot \delta\nabla S_{\theta,k}^n
    \right] \, .
\end{equation}
Here we have used Eqs.~\eqref{eq:deltax_kdef} and \eqref{eq:nablaS_kdef}, and we have additionally introduced
\begin{equation}
    \delta\nabla S_{\theta,k} = \nabla S_{\theta,k+1}-\nabla S_{\theta,k} \, .
\end{equation}
The superscripts $n$ in Eq.~\eqref{eq:costFunction-noLaplacian} indicate the trajectory along which $\nabla S_{\theta,k}$, $\delta {\bm x}_k$ and $\delta\nabla S_{\theta,k}$ are evaluated.

\end{document}